\RequirePackage{fix-cm}
\documentclass[twocolumn]{svjour3}          
\smartqed  
\usepackage{natbib}
\usepackage{graphicx}
\usepackage{xcolor}
\PassOptionsToPackage{hyphens}{url}\usepackage{hyperref}
\setcitestyle{aysep={}} 
\begin{document}

\title{A Simplified Gravitational Reference Sensor for Satellite Geodesy
}
%



\author{Anthony Dávila Álvarez$^1$ \and
        Aaron Knudtson$^1$ \and
        Unmil Patel$^1$ \and        
        Joseph Gleason$^1$ \and
        Harold Hollis$^1$ \and
        Jose Sanjuan$^1$ \and
        Neil Doughty$^2$ \and
        Glenn McDaniel$^2$ \and 
        Jennifer Lee$^2$ \and
        James Leitch$^2$ \and
        Stephen Bennett$^2$ \and
        Riccardo Bevilacqua$^3$ \and
        Guido Mueller$^1$ \and       
        Robert Spero$^4$  \and
        Brent Ware$^3$ \and
        Peter Wass$^1$ \and
        David Wiese$^4$  \and
        John Ziemer$^4$ \and
        John W. Conklin$^1$
}


\institute{John W. Conklin \at
    $^1$University of Florida, Gainesville, FL 32611 \\
    \email{jwconklin@ufl.edu}           
    \and
    \at
    $^2$Ball Aerospace, Broomfield, CO 80021
    \and
    \at
    $^3$Embry-Riddle Aeronautical University, Daytona Beach, FL 32114
    \and
    \at
    $^4$Caltech/Jet Propulsion Laboratory, Pasadena, CA 91109 \\
}

 \date{18 July 2021}

\maketitle

\begin{abstract}

The University of Florida, in collaboration with Caltech/JPL, Ball Aerospace, and Embry-Riddle Aeronautical University has developed a Simplified Gravitational Reference Sensor (S-GRS), an ultra-precise inertial sensor for future Earth geodesy missions. These sensors are used to measure or compensate for all non-gravitational accelerations of the host spacecraft so that they can be removed in the data analysis to recover spacecraft motion due to Earth’s gravity field, which is the main science observable. Low-low satellite-to-satellite tracking missions like GRACE-FO that utilize laser ranging interferometers are technologically limited by the acceleration noise performance of their electrostatic accelerometers and temporal aliasing associated with Earth’s dynamic gravity field. The current accelerometers, used in the GRACE and GRACE-FO missions have a limited sensitivity of $\sim10^{-10}$ m/s$^{2}$/Hz$^{1/2}$ around 1\,mHz. The S-GRS is estimated to be at least 40 times more sensitive than the GRACE accelerometers and 
over 500 times more sensitive if operated on a drag-compensated platform. The improved performance is enabled by increasing the mass of the sensor’s test mass, increasing the gap between the test mass and its electrode housing, removing the small grounding wire used in the GRACE accelerometers and replacing them with a UV LED-based charge management system. This level of improvement allows future missions to fully take advantage of the sensitivity of the GRACE-FO Laser Ranging Interferometer (LRI) in the gravity recovery analysis. The S-GRS concept is a simplified version of the flight-proven LISA Pathfinder (LPF) GRS. Performance estimates are based on models vetted during the LPF flight and the expected Earth orbiting spacecraft environment based on flight data from GRACE-FO. The relatively low volume ($\sim 10^4$\,cm$^3$), mass ($\sim$13\,kg), and power consumption ($\sim$20\,W) enables the use of the S-GRS on ESPA-class microsatellites, reducing launch costs and enabling larger numbers of satellite pairs to improve the temporal resolution of Earth gravity field maps. The S-GRS design and analysis, as well as its gravity recovery performance when implemented in two candidate mission architectures, are discussed in this article.
\keywords{Satellite Geodesy \and Gravity Measurements \and Accelerometers}
\end{abstract}

\section{Introduction}
\label{intro}

Precision inertial sensing is important for many space science missions, including fundamental physics experiments~\citep{everitt2011,touboul2001} and gravitational wave observation~\citep{danzmann2003}. The S-GRS presented here specifically targets future satellite geodesy missions, following the successful CHAMP \citep{reigber2002}, GRACE \citep{tapley2004}, GOCE \citep{drinkwater2003}, and GRACE-FO \citep{sheard2012} missions.

The GRACE (2002-2017) and GRACE-FO (2018-present) missions in particular have provided a nearly 20-year climate data record of Earth system mass-changes.  The utility of the data is incredibly diverse for serving both scientific and societal applications.  Contributions include quantifying the rate of ice sheet and glacier ablation globally, identifying and quantifying areas of unsustainable groundwater withdrawal due primarily to heavy irrigation, quantifying global and regional sea level changes and ocean heat content (when combined with satellite altimetry), monitoring drought severity through operational assimilation in the U.S. Drought Monitor, and assessment of geohazards, including mapping mass redistributions due to large earthquakes~\citep{tapley2019}. The importance of mass change observations was recognized in the 2018 Decadal Survey for Earth Science and Applications from Space, where Mass Change was listed as a Designated Observable~\citep{natacad2018}, and is now a core component of NASA’s Earth System Observatory to be launched in the next decade.

\begin{figure}[h!]
  \includegraphics[scale=0.52]{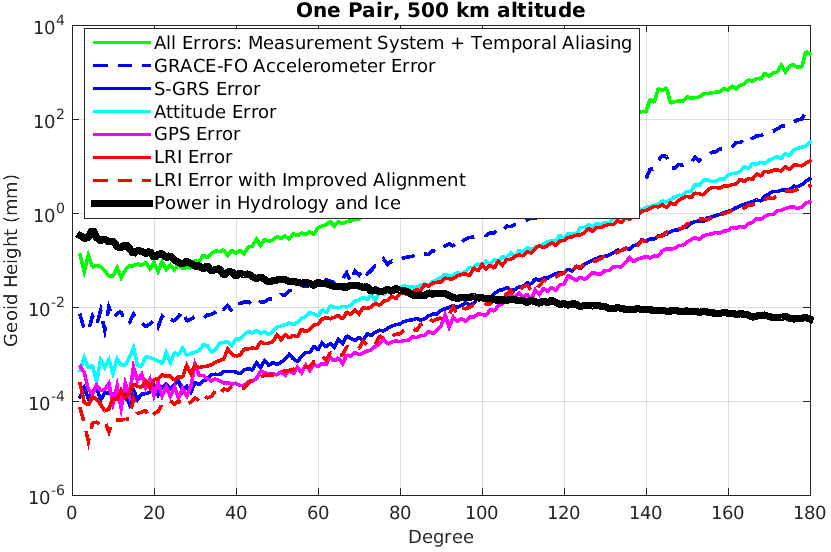}
  \includegraphics[scale=0.52]{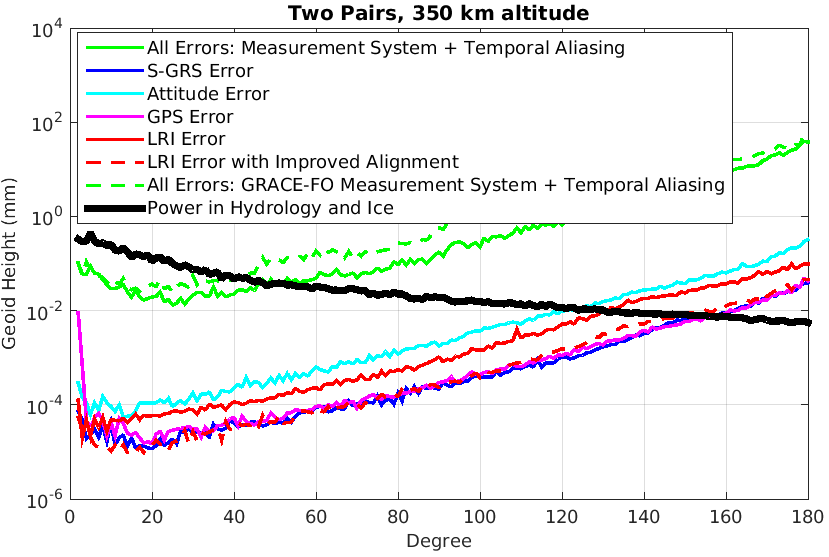}
\caption{Figure 1.  Monthly gravity field error budget (top) for a single pair of satellites at 500 km altitude in a polar orbit with no drag compensation and (bottom) an architecture consisting of two pairs (one polar, one inclined at 70 deg) at 350 km altitude with full drag compensation in the flight direction.  The error (units: mm of geoid height) of each instrument and its impact on the recovered gravity field per spherical harmonic degree is shown, as derived from  numerical simulations. Simulation details are provided in Appendix A}
\label{fig:degreeVar}       
\end{figure}

The quality of mass change data products derived from a gravity mission depends on the orbital characteristics of the mission (altitude, inclination, number of satellite pairs, etc.), as well as the precision of the onboard measurement system (inter-satellite ranging system, accelerometer, attitude and orbit determination). For GRACE-FO, it is well understood that the limiting source of error is due to temporal aliasing; i.e. undersampling of high frequency mass variations in the Earth system such as ocean tides and weather systems (Fig.~\ref{fig:degreeVar}, top). There are two possible ways to reduce the impact of temporal aliasing error on derived gravity fields: 1) sample more frequently via more satellite pairs~\citep{daras2018,elsaka2013,wiese2012,wiese2011a,wiese2011b} 
as shown in Fig.~\ref{fig:degreeVar} (bottom), and 2) process the data in an ‘along-the-orbital track’ approach rather than accumulating observations over a finite number of days, which is traditionally done.  Solution 1) is well understood, having been extensively studied in the literature, while solution 2) is less well understood but recent analysis suggests it is a promising approach. For example, first ‘along-the-orbital track’ analysis of the GRACE-FO LRI data revealed exquisite detail in static gravity field features otherwise not seen in traditional data analysis approaches~\citep{ghobadi2020,han2021a,han2021b}.

Future mission implementations should seek to minimize the impact of temporal aliasing error via one of the above two solutions. The precision of the accelerometer becomes important in such a design, as those in GRACE-FO limit both approaches. The precision of the GRACE-FO electrostatic accelerometers is insufficient for a multi-pair implementation (Fig.~\ref{fig:degreeVar} bottom. green dashed curve) at lower altitudes ($\sim$350\,km), and dominates over all other error terms. The S-GRS offers a technological pathway to improve the measurement of non-gravitational accelerations (Fig.~\ref{fig:degreeVar}, bottom) to an acceptable level for such constellation-type implementation. Furthermore, and perhaps even more importantly, the S-GRS offers a technological solution to measuring non-gravitational forces at a level that is on par with the precision of the LRI that was successfully demonstrated on GRACE-FO (Fig.~\ref{fig:degreeVar} shows comparable magnitude for the solid-blue and red-dashed curves)~\citep{abich2019}.  Such a match in performance between the accelerometer and the inter-satellite ranging instrument has potential to allow for full exploitation of the data in future ‘along-the-orbital track’ data analysis, as discussed in~\citep{spero2021}.  The current mismatch in performance between the LRI and the accelerometer on GRACE-FO limits interpretation of ‘along-the-orbital track’ signals to smaller spatial scales only; such analysis is currently not valid for the longest wavelengths in the gravity field, where the accelerometer error is dominant~\citep{ghobadi2018}, and the signal amplitudes are the largest.

\begin{figure}[h!]
  \includegraphics[scale=0.27]{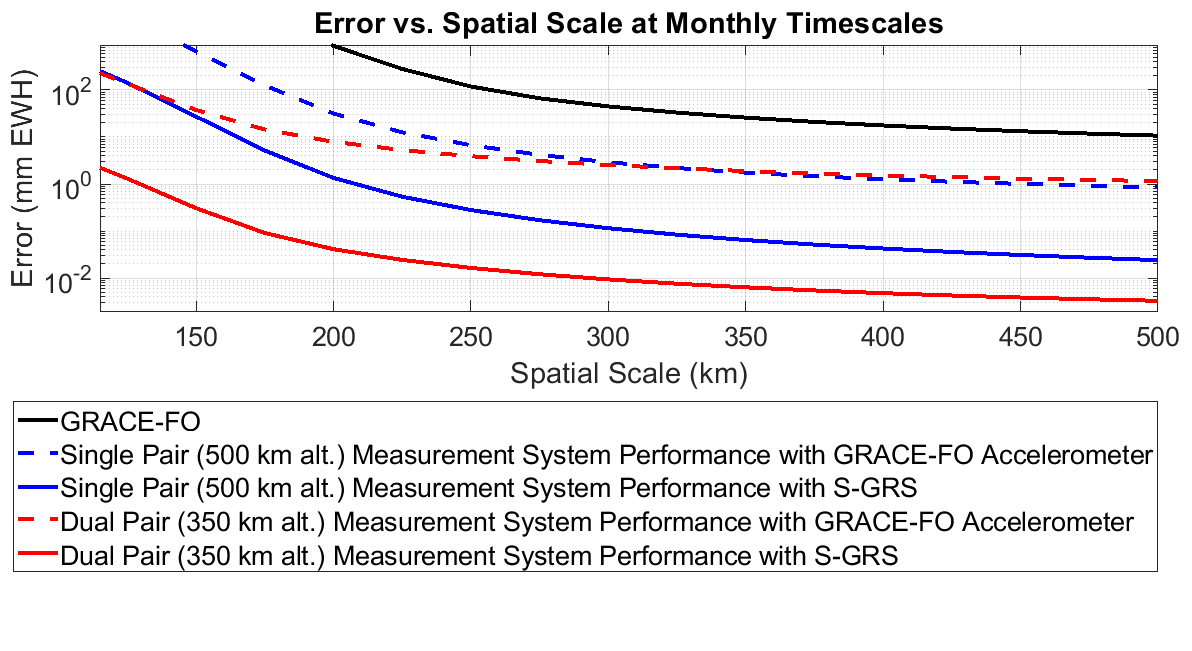}
\caption{Error in mm equivalent water height (EWH) versus spatial scale for monthly time resolutions, comparing measurement system performance with the S-GRS versus pre-flight error estimates from the GRACE-FO accelerometers as derived from numerical simulations. Note the blue and red curves neglect temporal aliasing error, showing the maximum potential improvement, while the black curve includes temporal aliasing error as well as instrument error models for the LRI, accelerometer, GPS, and attitude determination system, and as such is indicative of the performance of the GRACE-FO monthly gravity fields.}
\label{fig:spatialScale}       
\end{figure}

Figure~\ref{fig:spatialScale} uses methods described in~\citep{hauk2020} to show the maximum potential improvement the S-GRS offers in the architectures listed in Fig.~\ref{fig:degreeVar}. Maximum scientific potential can be assessed by looking at the measurement system performance (neglecting temporal aliasing error); this assumes temporal aliasing error is mitigated via ‘along-the-orbital track’ data analysis or future improvements in models of high frequency mass variations.  The potential for improvement offered by the S-GRS in the dual-pair architecture, in particular, could advance the measurement parameters  beyond those targeted in the Decadal Survey and identified in the Baseline Measurement Parameters in the Science and Applications Traceability Matrix identified by the Mass Change (MC) Study Team for the next observing system implementation~\citep{MC2020a,MC2020b}.

%
%
%

%
%

\subsection{Simplified Gravitational Reference Sensor}
%
The S-GRS design follows that of the flight-proven LPF Gravitational Reference Sensor (GRS) that represents the state of the art in precision inertial sensors~\citep{dolesi2003}. The LPF GRS used a 2\,kg, Au/Pt test mass (TM) inside a molybdenum electrode housing (EH). The housing held 12 gold-coated electrodes to differentially sense the position and orientation of the cube via capacitive sensing and to control it using electrostatic actuation. Six “injection electrodes” were driven with a 100 kHz AC bias voltage to frequency shift the capacitive measurement to high frequency. The 4\,mm gap (in LPF) between the TM and housing was defined by a trade-off; it was large enough to reduce noise sources, such as uncontrolled potentials on the electrodes, and small enough to measure TM displacement at the nanometer-level over the measurement bandwidth. A caging and venting mechanism used a series of mechanical fingers to secure the TM during launch and release it in orbit~\citep{bortoluzzi2012}. During science operations, the TM charge was controlled by a charge management system (CMS) based on UV photoemission using Hg vapor lamps as the UV source~\citep{wass2018}. The CMS eliminates the need for the small grounding wire used in the ONERA accelerometers that both limits their performance and causes challenges during integration and testing~\citep{touboul1999}. 

The primary performance metric for these instruments is the acceleration noise from the residual spurious forces acting on the TM, often defined by an amplitude spectral density (ASD). 
LPF, launched in 2015, exceeded requirements by more than a factor of ten with a measured acceleration noise below 3$\times10^{-15}$\,m/s$^{2}/{\rm Hz}^{1/2}$ between\,0.4 mHz and 20\,mHz~\citep{armano2018}, the same frequency band that is important for Earth geodesy. This represents a factor of $3\times10^{4}$ improvement over the performance of GRACE and GRACE-FO and 10$^{3}$ over that of the GOCE instrument \citep{christophe2010}. 

The S-GRS is a scaled-down version of the LPF GRS, with reduced mass and complexity, and it is optimized with respect to performance for future geodesy missions (see Fig.~\ref{fig:degreeVar}). Key features of the design and its implementation that improve sensitivity with minimal size, mass, and power are:

\begin{itemize}
    \item The TM grounding wire used in previous accelerometers is removed, TM charge is instead controlled via non-contact UV photoemission as in LPF. The grounding wire is one primary source of acceleration noise in current electrostatic accelerometers.
    \item Removing this grounding wire allows us to increase the TM-to-EH 
    gap from $\sim$100\,$\mu$m to $\sim$1\,mm, as well as the TM mass from $\sim$100\,g to $>$500\,g. 
    Acceleration noise (for surface forces) scales inversely with TM size and density, and to first order, with the gap size between the TM and EH.
    \item Careful control of the environment surrounding the TM, including venting to space to reduce residual pressure, and shielding against Earth’s magnetic field and thermal fluctuations.
    \item Tightly integrating the S-GRS and LRI, eliminating the microwave ranging instrument used on GRACE-FO to minimize size, mass, and power.
    \item Design for the possibility of operating the S-GRS on a drag-compensated platform to further reduce acceleration noise, while simultaneously maintaining orbit altitude. 
\end{itemize}

\begin{figure}[h!]
  \includegraphics[scale=0.15]{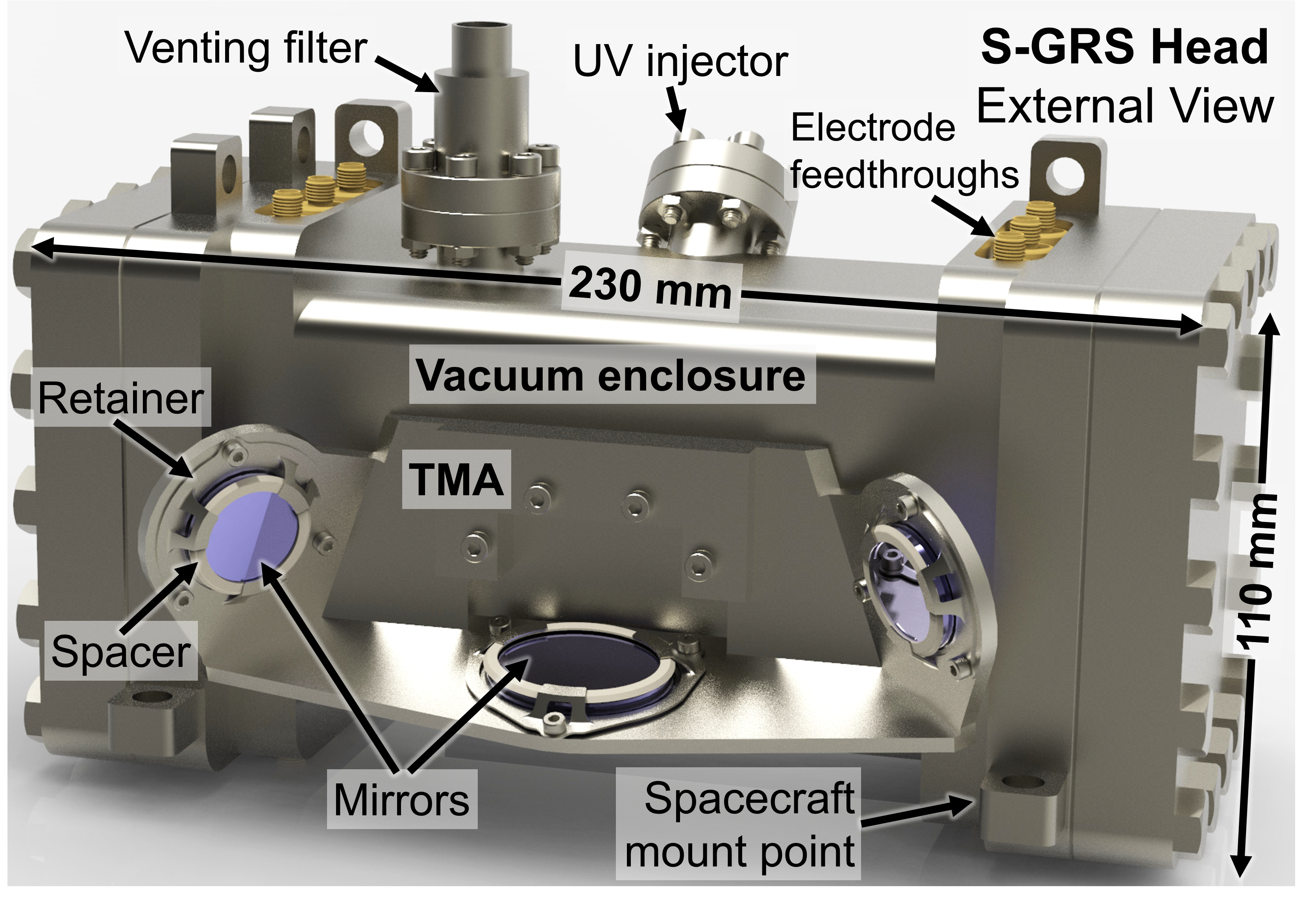}
  \includegraphics[scale=0.13]{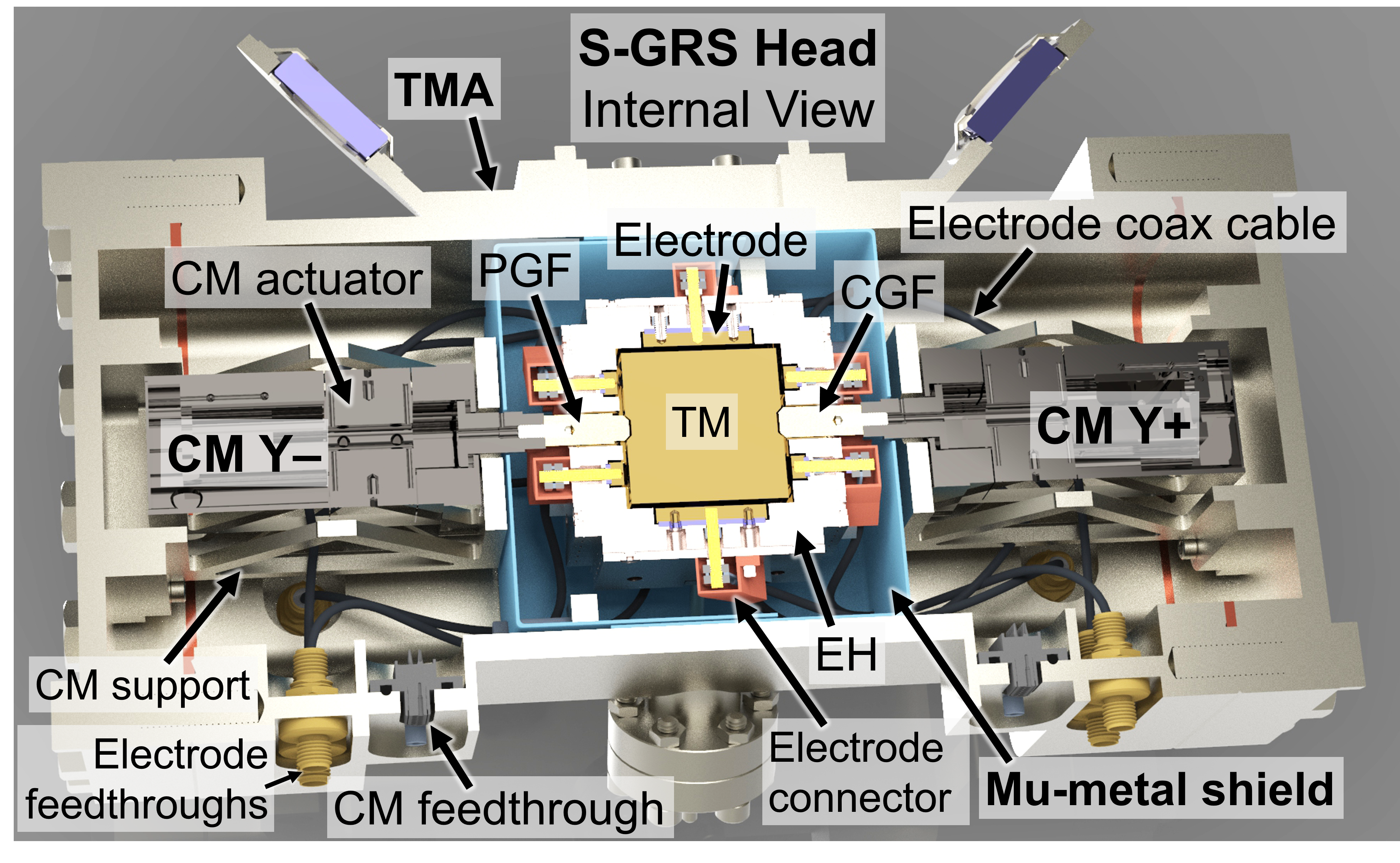}
\caption{S-GRS Head CAD rendering external view (top) and internal view (bottom). CM: Caging Mechanism, PGF: Pyramidal Grabbing Finger, CGF: Conical Grabbing Finger, TM: Test Mass, EH: Electrode Housing, TMA: Triple Mirror Assembly.}
\label{fig:S-GRS_Head}       
\end{figure}

The instrument consists of an S-GRS Head and an Electronics Unit. The detailed CAD design of the S-GRS Head is shown in Fig.~\ref{fig:S-GRS_Head}. It is the mechanical unit containing the TM and EH, the caging mechanism, and the vacuum enclosure with a compact version of the triple mirror assembly used on GRACE-FO. The Electronics Unit performs all control and readout functionalities and is separated from the S-GRS head to minimize electromagnetic and thermal disturbances to the TM. In the remaining sections of the paper each sub-element of the S-GRS will be described beginning with the S-GRS head in Section 2 and the S-GRS electronics unit in Section 3. In Section 4 we present a breakdown of the S-GRS acceleration noise performance followed by simulations of the test-mass position control algorithms in Section 5 and conclusions in Section 6.

\section{S-GRS Head}

\subsection{Test mass and electrode housing}

\paragraph{}
Detailed performance analyses determined that a 30 mm cubic Au-Pt TM with a mass of 540\,g results in the desired acceleration noise performance described in Section~\ref{sec:performance}, while keeping mass and volume of the sensor to a minimum. The TM is a gold coated cube with two indented features on opposite sides, one conical and the other pyramidal, used to hold the test mass on ground and during launch.

\begin{figure}[h!]
  \includegraphics[scale=0.7]{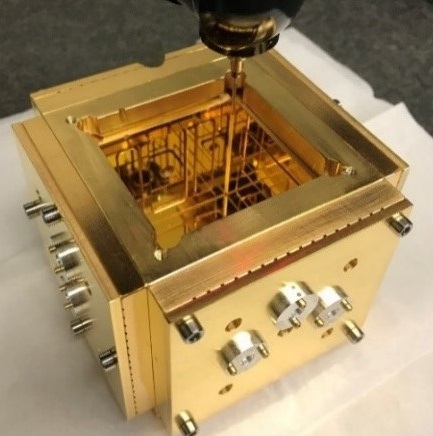}
  \includegraphics[scale=0.7]{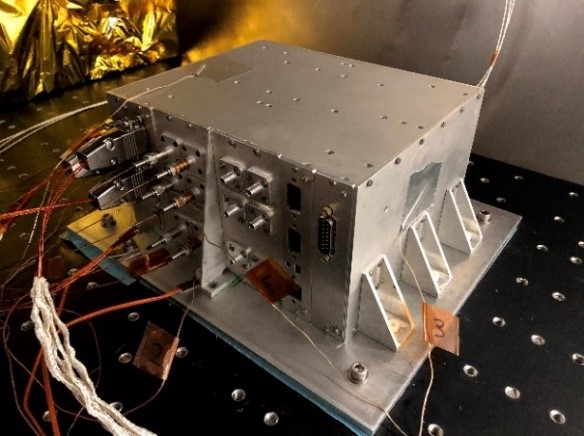}
\caption{LISA-like EH prototype (left) and TRL 5 LISA CMS during thermal-vacuum (TVAC) testing at UF (right). The S-GRS EH and Electronics Unit are based on these designs.}
\label{fig:EH_CMS}       
\end{figure}

The electrode housing design is a scaled-down version of the LISA-like EH prototype previously fabricated by the UF team and shown in Fig.~\ref{fig:EH_CMS}. The inner dimension of the cubic EH is 32\,mm creating a 1\,mm gap between the TM and EH, and its wall thickness is approximately 10\,mm. The technology readiness level (TRL) 5 EH will be fabricated from gold-coated molybdenum with sapphire electrode spacers. Molybdenum is chosen for its low magnetic properties, machinability, and high thermal conductivity, which reduces thermal gradients across the sensor. Sapphire is chosen because its coefficient of thermal expansion is nearly the same as that of molybdenum.

\begin{figure}[h!]
  \includegraphics[scale=0.2]{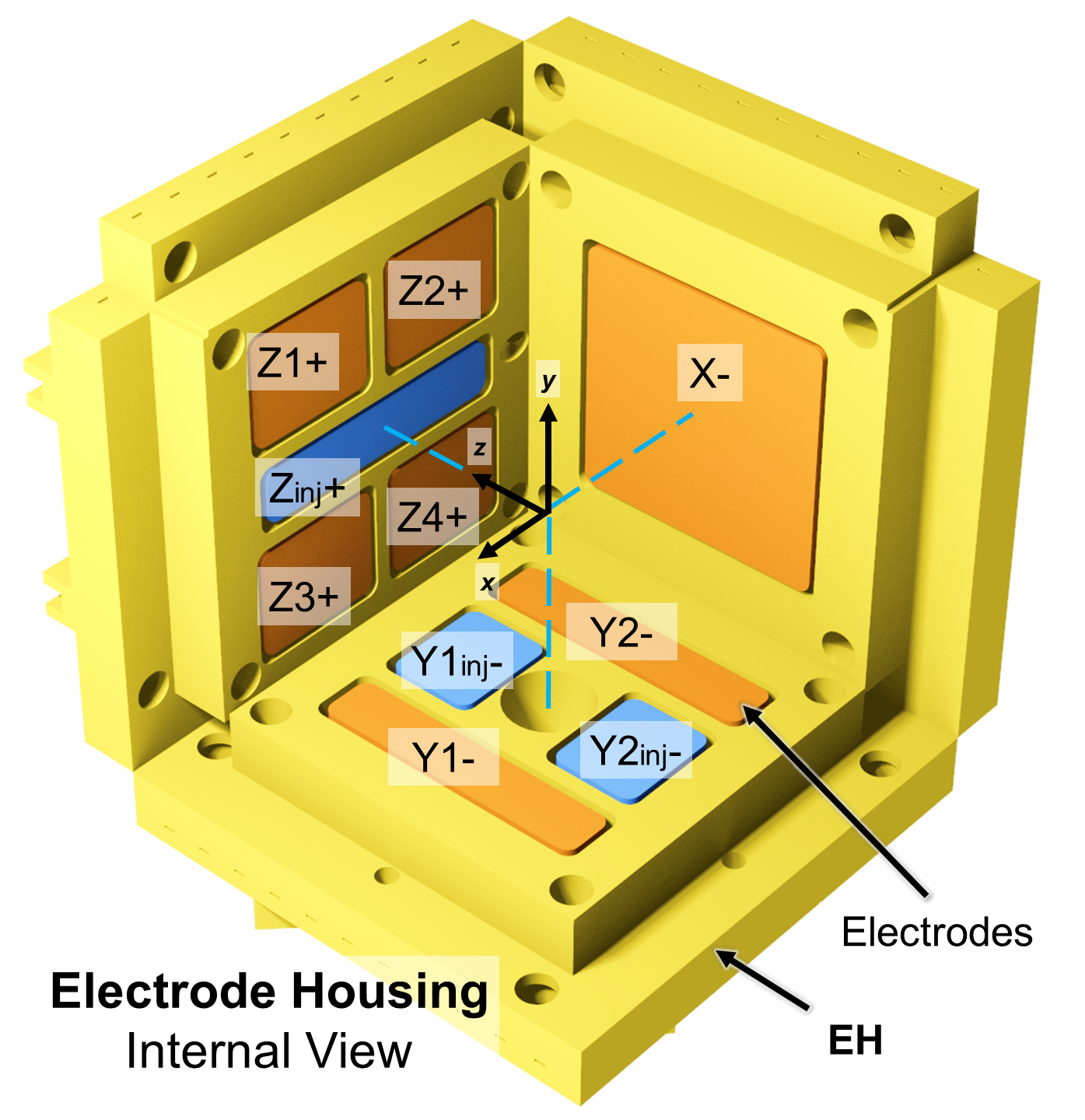}
  \caption{Internal view of the EH with labeled electrodes. Blue electrodes are connected to the injection signal, while orange ones are used for sensing and actuation.}
  \label{fig:EH}
\end{figure}

The electrode geometry, shown in Fig.~\ref{fig:EH}, differs from that of LPF. The S-GRS design has only one large electrode to sense the test mass motion in the sensitive direction. This maximizes sensitivity in this direction and minimizes cross coupling with other degrees of freedom. Around the holes used for the CM fingers, both faces normal to the Y-axis have 2 injection signal electrodes and 2 sensing electrodes that can measure TM motion along the Y-axis and rotation along the Z-axis. Faces normal to the Z-axis have 1 injection electrode and 4 sensing electrodes to detect TM motion along the Z-axis and rotation along the X and Y axes. 


\subsection{Triple Mirror Assembly}
\label{sec:TMA}

\begin{figure}[h!]
  \includegraphics[scale=0.11]{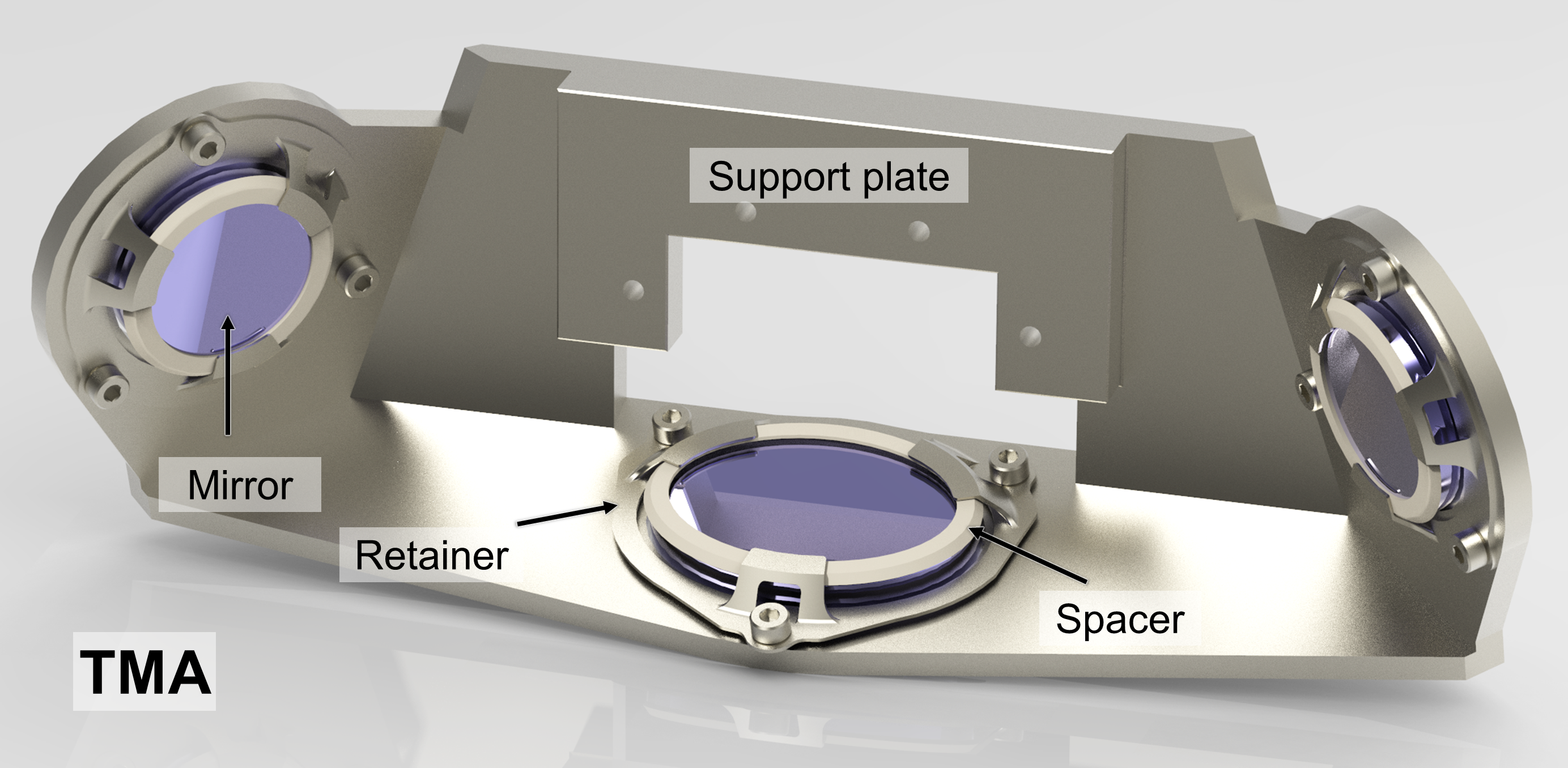}
  \caption{Close-up of the TMA and its components.}
  \label{fig:TMA}
\end{figure}

The Triple Mirror Assembly (TMA), shown in Fig.\,\ref{fig:TMA}. is an optics mounting structure located on the +X face of the vacuum enclosure (see Section \ref{sec:VE}). It acts as a retroreflector at the ends of the “racetrack” configuration for the LRI that measures the distance between the pair of satellites~ \citep{dahl2016}. The input beam is reflected in three orthogonal planes, producing an output beam at an offset distance in an anti-parallel orientation. The TMA is designed so that the virtual vertex at which the three planes intersect is coincident with the center of mass of the spacecraft’s inertial reference body, the TM, when it is located in the center of the EH. This way, the LRI’s distance measurement is done between the reference masses of each satellite.

The concept for the TMA is based on an equivalent component of the GRACE-FO mission~\citep{abich2015,GRACE-FO}. The corner-cube principle is similarly implemented in both designs, with the purpose of returning the LRI beam, regardless of pointing variations. The proposed TMA is significantly smaller than the one used in GRACE-FO. This is because in GRACE-FO, the laser interferometer had to be built around a Ka-band microwave link which was used as the primary interferometer.

The TMA is made of titanium alloy, with three sapphire mirrors supported by titanium alloy retainers and PEEK spacers. 
It has a yaw angle of 45$^{\rm o}$ from the chamber wall for a symmetric aperture of the mirrors in the input and output planes. It also has a pitch angle of 20$^{\rm o}$ that reduces the volume of the body that supports the mirrors. There is a mean distance of 129 \,mm between the input and output beams.



An advantage of the symmetric geometry of the TMA and its large contact area with the vacuum chamber wall of the same material (see Section \ref{sec:VE}), is the athermal behavior on the vertex position with respect to the TM center. Any thermal deformation of the chamber would similarly affect the TMA planes, reducing any changes in the vertex position. 
The goal is to maintain the TMA vertex within 50 $\mu$m of the TM center and a co-alignment error of less than 50 $\mu$rad.

\subsection{Vacuum Enclosure}
\label{sec:VE}

The S-GRS electode housing and TM are enclosed within a vacuum enclosure, shown in Figure \ref{fig:S-GRS_Head} to limit the effects of Brownian noise and improve the sensing system’s noise environment \citep{mcnamara2013} with a goal pressure of $10^{-5}$~Pa. While on ground, the enclosure can be pumped out for testing purposes. On orbit, residual gas in the chamber will be vented to space through a duct to separate the system from the outgassing of the rest of the vehicle. The chamber is also the main supporting structure; the TM is located at its center, the upper and lower sections of the caging mechanism are positioned on opposite sides of the EH, and the TMA is mounted on the external +X face of the enclosure (see Section \ref{sec:TMA}).

The vacuum enclosure is designed to support the TMA on its exterior and contain the TM, EH, caging mechanism, magnetic shield, and their supports. The EH is fastened and aligned through its frame to stiffener-like surfaces within the chamber that are placed away from the TM center. These surfaces are also used to support the standoffs and fasteners for the magnetic shield that surrounds the EH. Twenty coaxial SMA feedthroughs that connect to the EH electrodes are distributed around the top and bottom sections of the enclosure, except on the TMA face. Two D-Sub 25 pin feedthroughs that connect to the caging mechanism actuators are placed on the -X face. The UV injector is attached to a CF flange located at the central section of the enclosure between the +Z and -X faces at a 45 deg angle along the +Y axis. The enclosure enclosure vent duct is attached to a flange on the +Z face.

The chamber is based on the LPF design but with a smaller size and rectangular shape. The 5~mm thick walls were selected to keep the maximum deformation from pressure and thermal loads below 10~$\mu$m. With the ConFlat-type (CF) square ends, the chamber is 235~mm long with an internal square section of 78~mm at the center.

The UV injector feedthrough and flange are based on commercially available CF options, similar to the one used in the LISA Inertial Sensor UV Kit (ISUK) \citep{wass2018}. The 65 mm injector is supported at the flange and extends through a hole on the EH. The diameter and length of the injector is such that no additional support is required to avoid any collisions to the EH during launch. The central location along the Y axis is required so that the UV light is reflected at a 45 deg angle of incidence between the TM and the +Z injection electrode of the EH. 

In addition to the protective walls of the vacuum enclosure, a magnetic shield is required to attenuate the magnetic field at the TM (see Fig.~\ref{fig:S-GRS_Head}). The cubic shield 
fully envelops the sensor with the exception of gaps for the support stiffeners and holes for the caging mechanism grabbing fingers (GF), coaxial cables, UV injector, and shield fasteners. 
Acceleration noise modeling described in Section~\ref{sec:performance} shows that attenuation of the Earth’s magnetic field by at least a factor of ten is needed to meet performance goals.  

%

\subsection{Caging Mechanism}

\begin{figure}[h!]
  \includegraphics[scale=0.18]{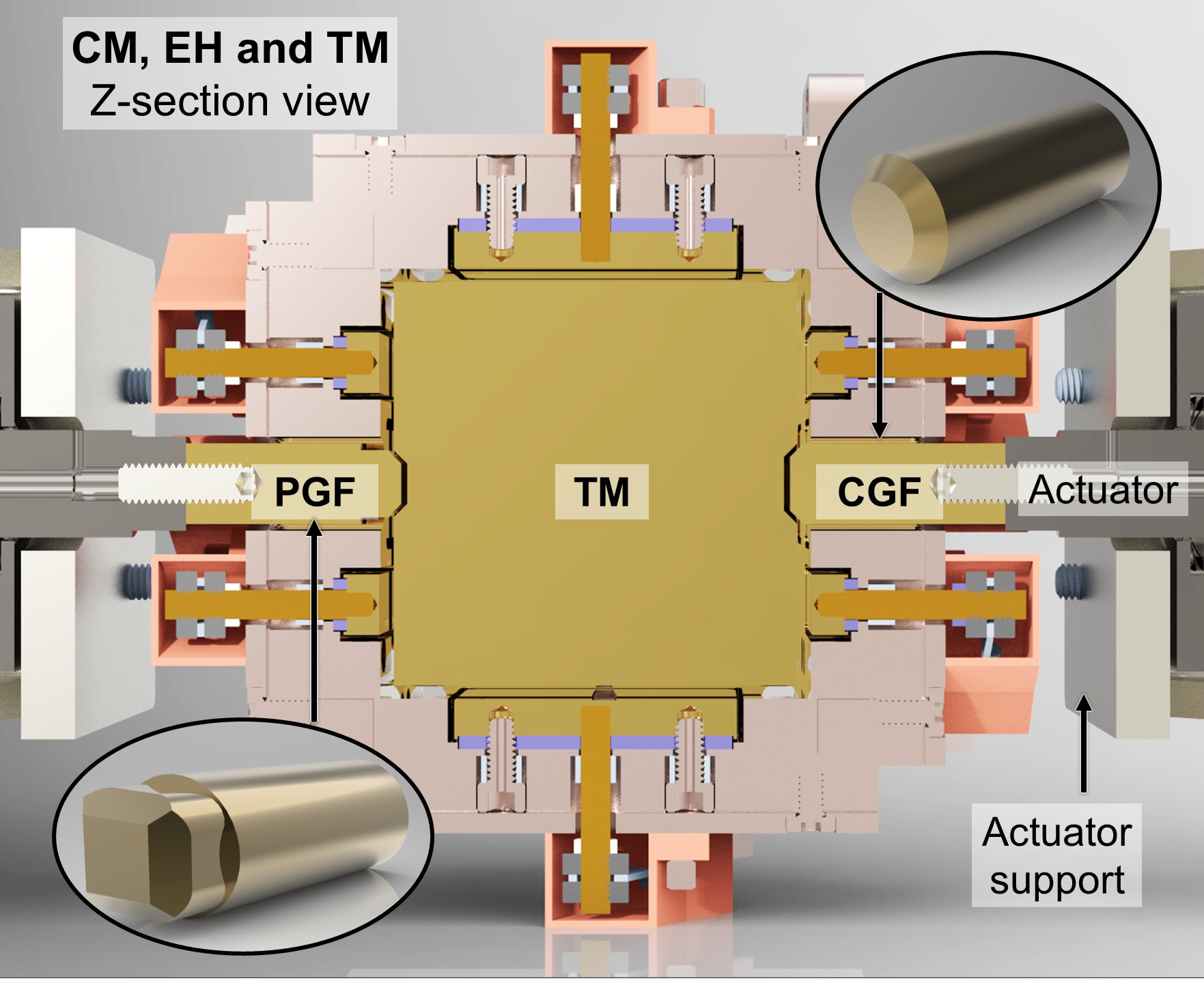}
  \caption{Section view of the CM around the EH and TM. The shapes of the PGF and CGF tips are shown.}
  \label{fig:CMview}
\end{figure}

The caging mechanism must secure the TM against launch loads and gently release it once in orbit. The technical challenge is that upon release, the TM must have a low enough velocity relative to the spacecraft so that the S-GRS control system can electrostatically capture it. To achieve such a low relative velocity, the design must minimize both adhesion between the GF and TM gold surfaces, and any asymmetry in the mechanical preloads just prior to release \citep{bortoluzzi2012}. 
The S-GRS caging mechanism reduces the three-stage LISA system to a simpler two-stage system. The first stage uses standard aerospace launch lock actuators to handle launch loads, while the second stage utilizes a commercial vacuum-compatible piezo-walker motor. The first stage applies ~200\,N to a pair of opposing titanium alloy, hard gold plated fingers that engage the TM at the conical and pyramidal indents, as seen in Fig.~\ref{fig:CMview}.

\begin{figure}[h!]
  \includegraphics[scale=0.17]{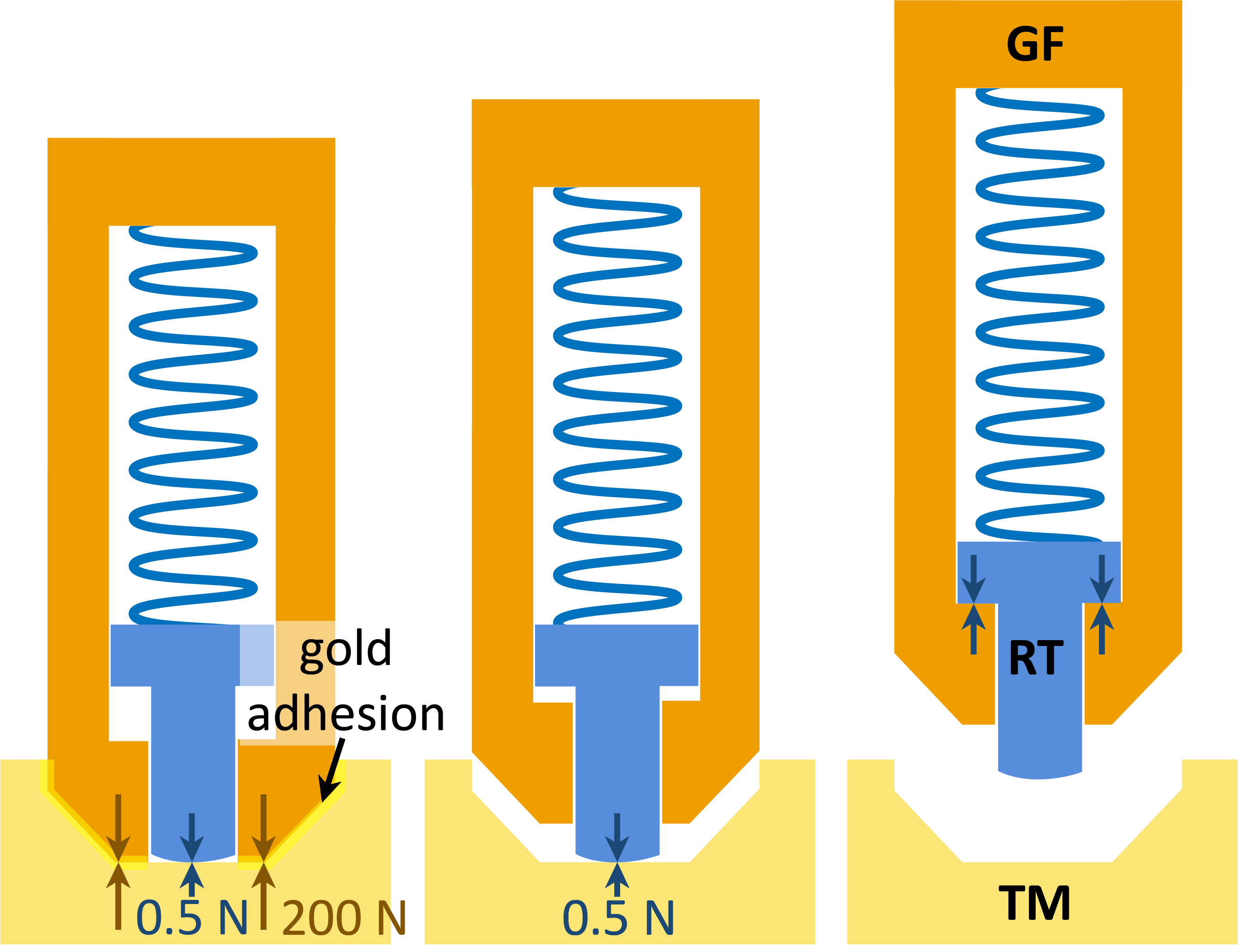}
  \centering 
  \caption{Spring-tip concept release process. First, the GF applies a 200\,N to the TM, As the GF retracts (center), a 0.5 N preload applied by the RT breaks the gold adhesion. Finally, the RTs damp the TM motion until they are fully retracted. The PGF and CGF have identical RTs.}
  \label{fig:CMrelease}
\end{figure}

After the one-time first stage launch lock is released, the second stage retracts the gold alloy fingers until the TM is held with a $<$1\,N preload by spring-loaded release tips (RT), as seen in Fig.~\ref{fig:CMrelease}. The spring stiffness is designed to ensure that the adhesion between the gold-coated TM and the gold alloy fingers is broken. The final RTs are fabricated from a hard, non-stick silicon nitride material to avoid adhesion with the TM. Damping is added to the system so that the TM motion is minimized upon final retraction and release. To minimize potential electrostatic interactions between the silicon nitride RTs and the TM, the caging mechanism retracts the tips $\sim$10\,mm inside the EH.

The simplification to a two-stage system is possible because of the lighter 0.54\,kg mass of the S-GRS TM compared to the 1.96\,kg LPF TM. Static and random vibration FEA simulations validate the use of only two fingers to support the TM during launch. The static simulations apply a 15\,$g$ acceleration to the system in three orthogonal directions. The random vibration simulations use the 14.1\,$g_{\mathrm{rms}}$ ASD from the NASA GEVS document \citep{GEVS}. 
The result proves the feasibility of the two-finger approach with a high margin of safety.
The 200\,N holding force safely satisfies the low TM displacement requirements, especially for the critical case of random vibration along the Y-axis (along the length of the fingers).

In addition to its caging capabilities, the mechanism must also release the TM at a low residual velocity. A maximum allowable residual velocity is obtained from a 2 DOF dynamics simulation that starts immmediately after the TM is released and stops when it has been captured by the electrostatic actuation system or collides with the EH. The simulation iterates through a range of initial TM linear and angular velocities, actuation voltages, and directions of motion. The objective is to find the highest initial velocity values that can still be captured by the actuator system. In the model, the TM is treated as a square of four rigidly grouped particles that represent its four corners in the plane of motion. As the square moves because of its initial velocity, the actuator system applies a force that attempts to cancel the TM motion. The electrostatic force is a non-linear function, dependent on the voltage and distance between the TM and actuator electrodes, and the electrode area \citep{ciani2008}. Since the S-GRS has different electrode configurations on each face, the electrode areas and locations vary for each plane of motion. The greater the electrode area and the smaller the gap to the TM, the higher the actuation force. 
Assuming a 10 V actuation voltage, the residual linear velocity must be $<$4.5 $\mu$m/s and the angular residual velocity $<$300 $\mu$rad/s.

With a known residual velocity requirement, a second dynamics model is created to simulate the actual retraction and release process. The 1 DOF model is based on that used for the LPF Grabbing Positioning and Release Mechanism (GPRM) \citep{bortoluzzi2012} but with the Ball Aerospace spring-tip concept. The system uses the retraction speed of both GFs as an input and simulates the RT and TM dynamics. Assuming all conditions are symmetric between both sections of the caging mechanism, the result is an ideal release where the TM remains centered in the EH. It is only when asymmetries are included in the system, that the TM drifts to one direction. The worst-case scenario for this asymmetry involves having the maximum adhesion force profile on one GF and no adhesion with a retraction delay on the other. The retraction asymmetry causes a kick from the delayed GF, which acts like a compressed spring \citep{bortoluzzi2016}. The one-sided adhesion force keeps the TM attached to only one GF, pulling the TM as it is retracted. This requires assuming that the gold adhesion force profiles described in the GPRM model apply between the GF and TM. Other than the mechanical spring and damping forces between the GFs and their respective RTs, all bodies in contact are treated as compressed springs to simulate contact forces with the penalty method \citep{fourment1999}. An additional electrostatic spring-like force acts upon the TM and is a function of the squared sensing injection bias \citep{ciani2008}. 

\begin{figure}[h!]
  \includegraphics[scale=0.23]{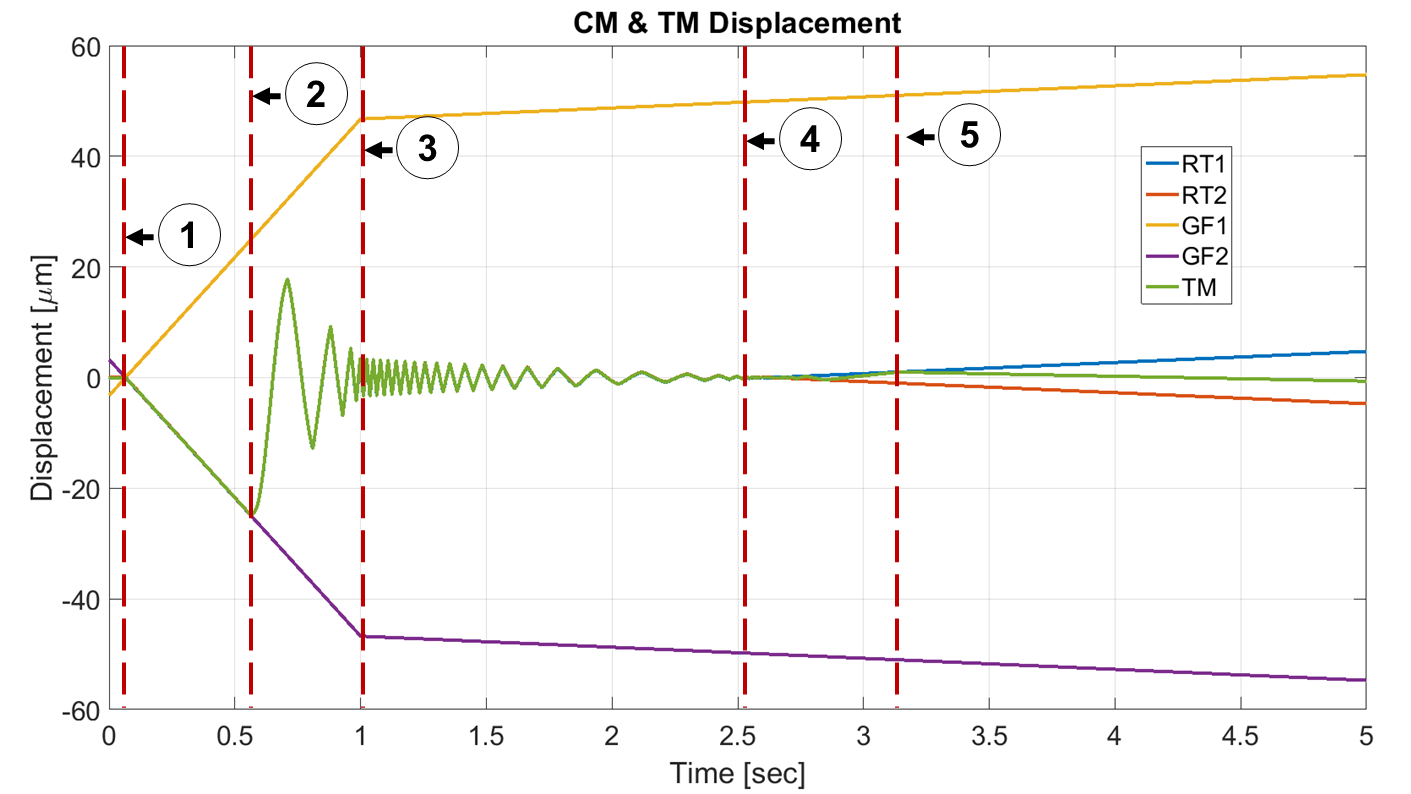}
  \includegraphics[scale=0.23]{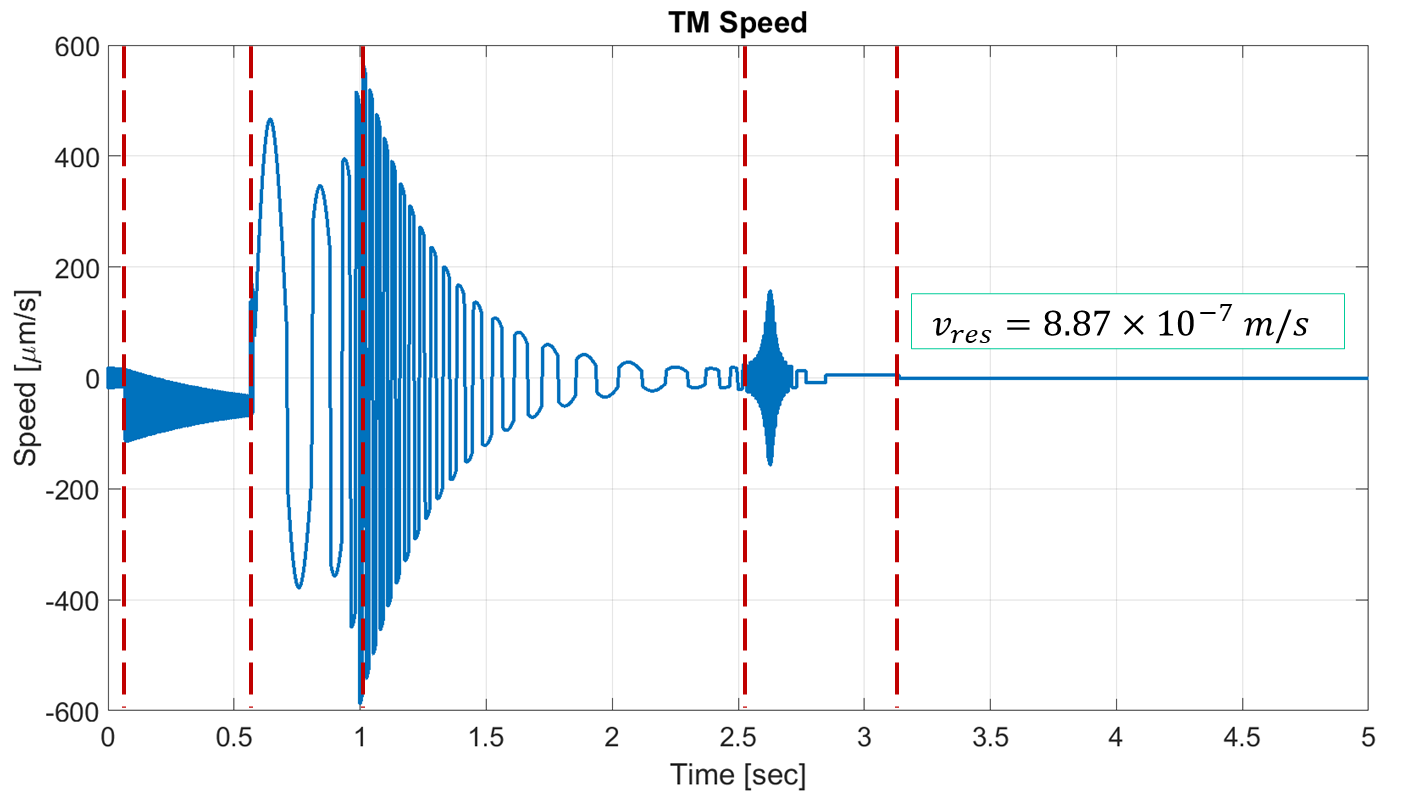}
  \caption{Time series of the spring-tip concept release simulation. The top plot shows the displacement of each component along with labels for the described events. The TM speed oscillates as its motion is dissipated by the RTs bottom).} 
  \label{fig:rel_sim}
\end{figure}

The model’s output contains the location, velocity, and forces of each component along the Y-axis (direction of release). These can be used to identify the necessary spring and damping coefficients, retraction velocity, and understand the performance of the current design. The spring-tip concept worst-case simulation uses the 200 N preload, a 50 $\mu$s delay (as used in the GPRM model), and the strongest adhesion force profile on one GF. For the resulting time series on Figure \ref{fig:rel_sim}, a damping ratio of 3, a spring constant of 131 N/m, an initial actuator speed of 50 $\mu$m/s, and an injection signal bias of 4 V is used. In the figure, a time series describes the displacement, speed and the GF to TM contact forces. Initially, the GFs are preloaded against the TM until retraction starts. Because of the adhesion force, the TM sticks to GF2 as it is retracted (seen in point 1 of the top plot). In point 2, RT1 is pulled back by GF1 as it is constrained by its geometry inside the GF. This decreases RT1's preload to the TM, allowing RT2 to push the TM and break the adhesion force. With no contact between the GFs and TM, the retraction speed is decreased at point 3 to extend the time in which the RTs dissipate the TM kinetic energy. As seen in the middle plot, the TM speed amplitude decays while GF retraction continues. At point 4, both RTs are constrained and pulled by their respective GFs until the TM is fully released in point 5. 

\begin{figure}[h!]
  \includegraphics[scale=0.2]{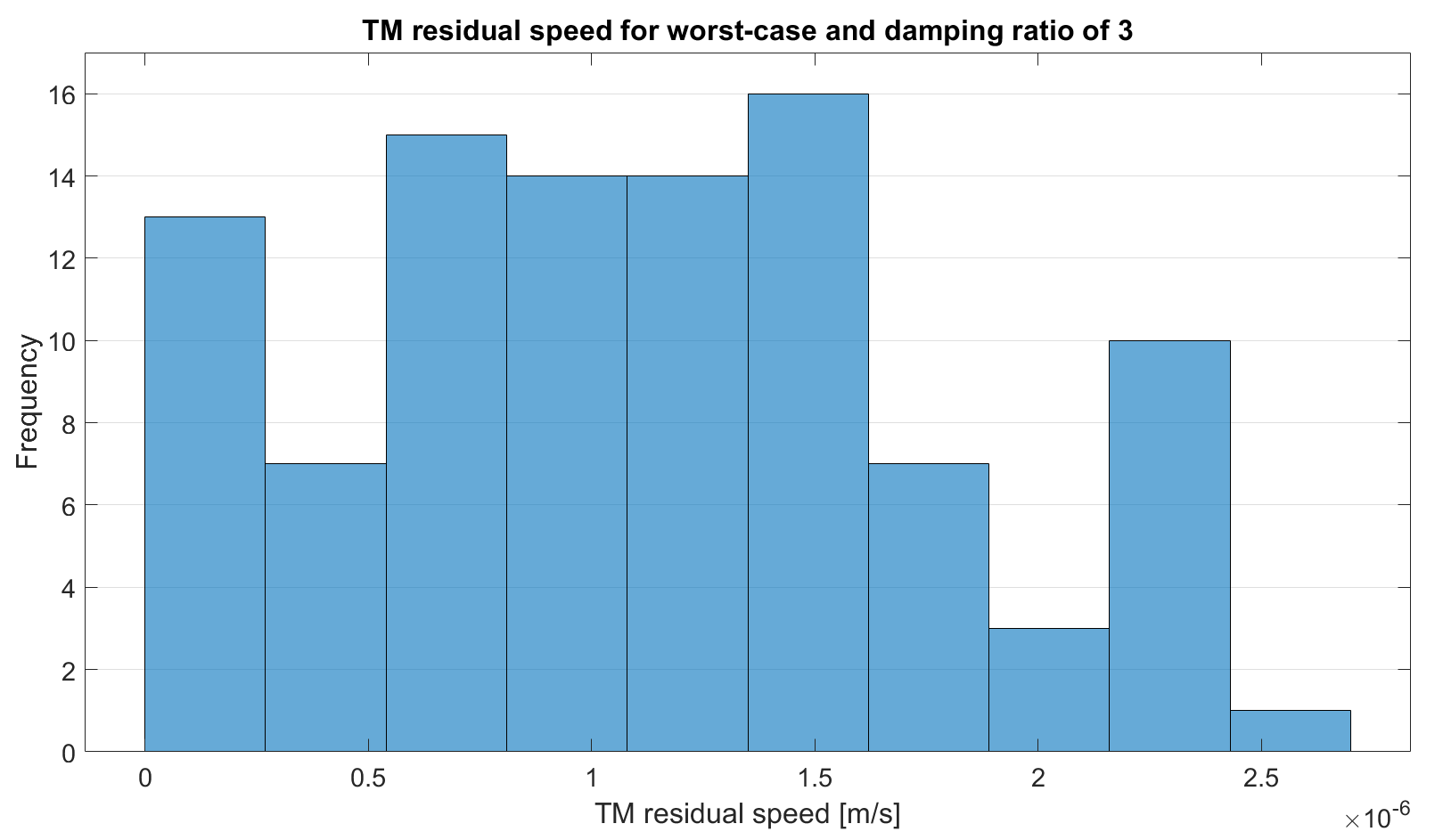}
  \caption{Histogram of TM residual speed from a 100 point Monte Carlo analysis.}
  \centering
  \label{fig:rel_hist}
\end{figure}

The adhesion force profile and retraction delay time would be random and uncontrollable inputs in the actual release process, so a Monte Carlo simulation is used to determine the probability of safe release. The adhesion force maxima ranges from 0.1 to 0.15 N. The elongation or separation between the bonded surfaces in which the maximum force occurs varies from 0.4 to 0.8 $\mu$m. The retraction delay time ranges from 0 to 100 $\mu$s. Using a damping ratio of 3 and assuming all three random inputs are uniformly distributed, results in a mean TM residual velocity of 1.11 $\mu$m/s with a standard deviation of 0.67 $\mu$m/s for 100 points. A histogram of the results can be seen in Fig.~\ref{fig:rel_hist}, where the TM is safely released for all points with a minimum safety factor of 1.7 in the right-most bar. A multiple DOF model is being developed to estimate the TM angular velocity distribution.


\subsection{Charge Management System}

The University of Florida team has been developing the LISA CMS for over four years~\citep{kenyon2021}. The CMS utilizes newer UV LEDs that both resolve technical issues related to the Hg-lamp system used by LPF, and reduce the size, weight and power of this subsystem. Compared to the Hg lamps used in LPF, UV LEDs are smaller, lighter, consume less power, and have a higher dynamic range, with at least an order of magnitude improvement in each performance area~\citep{olatunde2015}. The S-GRS will use an advanced pulsed, continuous charge control scheme with a low power (nW) UV light, pulsed in a low duty cycle that is synchronized to the 100 kHz injection electric field~\citep{inchauspe2020}. Photoelectron flow between the TM and EH will anti-align with the injected field, causing the phase of the UV light pulses with respect to the field to determine the net flow of photoelectrons and the resulting equilibrium charge of the TM. The UV pulse phase is then selected to produce a passively stable TM equilibrium charge of 0 Coulombs.

The CMS unit, shown in Figure \ref{fig:EH_CMS}, consists of an FPGA-based controller, a power system, the UV LEDs and their current sources, a mechanical enclosure, and a fiber optic harness that delivers the UV light to the S-GRS head. The LISA CMS will reach TRL 6 by the end of 2023. In terms of S-GRS development, the CMS only needs to be tested in integration with the TM and EH because of the difference in geometry to the LISA GRS. 
In addition to being smaller in size, the S-GRS will only have a single UV light feedthrough (instead of three for LISA) that will be oriented to maximize discharge robustness against variations of the gold surfaces of the TM and EH.

\section{Electronics Unit}

The S-GRS electronics, which comprises both analog and digital elements, will perform the following functions:
\begin{itemize}
    \item Generate (digitally) the TM injection bias to frequency shift the capacitive readout 
    to the injection frequency (100 kHz)
    \item Perform the six degree of freedom position readout of the TM by differencing opposing pairs of sensing/actuation electrodes via analog electronics, and demodulating the differenced signals at the injection frequency via digital electronics
    \item Apply actuation voltages via a digital-to-analog converter to the sensing and actuation electrodes at audio frequencies to control the position and orientation of the TM
    \item Drive the current source for the UV LEDs to discharge the TM
    \item Drive the caging and positioning mechanism
\end{itemize}

The S-GRS can be operated in two primary modes: drag-compensation and accelerometer mode. In accelerometer mode, all control algorithms are embedded within the S-GRS so that it may be treated as a black box that provides spacecraft acceleration estimates. The TM position and orientation is estimated and a control algorithm commands the actuation system to keep the TM centered in its housing. The command forces applied to the TM and the residual spacecraft to TM motion are used to estimate the non-gravitational spacecraft acceleration. In this mode, no spacecraft propulsion is required. 

Drag-compensation is a spacecraft-level control loop that commands thrusters using S-GRS data to keep the spacecraft centered on the TM without applying actuation forces to the TM. In this mode, external forces on the spacecraft are directly canceled by a propulsion system, minimizing the applied forces and therefore the force noise on the TM. In addition to improved acceleration noise performance, drag-compensation also simplifies the data analysis, since there is no non-gravitational spacecraft acceleration that must be accounted for. For an infinite-gain closed-loop system, the TM position signal is zero but the noise levels of the measurement will limit the performance of the loop and, consequently, the ability to measure non-inertial forces in accelerometer mode or minimize them in the drag-free configuration.

\subsection{Capacitive Sensing Electronics}
The TM position in the EH is measured by the capacitive sensing electronics. The basic principle consists of measuring the differential changes between the capacitances on opposite sides of the TM. Similarly, the rotation of the TM can be measured by combining the information of electrode pairs.

\begin{figure}[h!]
\centering
\includegraphics[width=\linewidth]{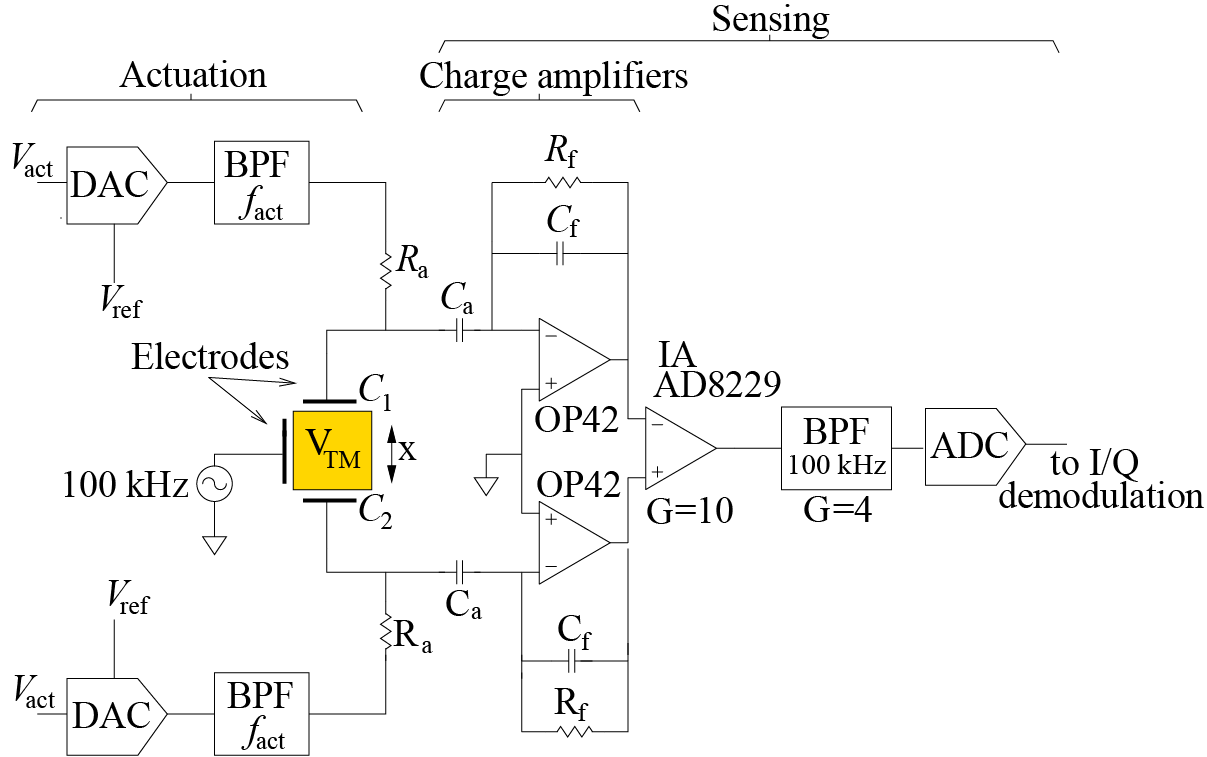}
\caption{Capacitive sensing and actuation electronics for one degree of freedom. Two charge amplifiers (with $C_{\rm f}$=10\,pF and $R_{\rm f}$=1\,M$\Omega$) 
and an IA 
are used to measure the differential change in the capacitances $C_{1}$ and $C_{2}$ due to TM motion along $x$. A multi-feedback band-pass filter after the IA serves as an anti-aliasing filter and adds extra gain to minimize ADC quantization noise. The components of the design are chosen such that they have space qualified equivalent versions. Actuation signals are applied at $\sim$100\,Hz through a DAC and a band-pass filter. $R_{\rm a}$=1\,M$\Omega$ and $C_{a}$=1\,nF decouple the sensing and actuating signals. \label{fig.elec.1}}
\end{figure}

One of the limiting noise sources is the intrinsic noise in the capacitive sensing electronics. The main goal is to reach noise levels on par with the GRACE-FO LRI system, i.e., around 1\,nm/${\rm Hz}^{1/2}$ in the frequency range of approximately 0.35\,mHz to 50\,mHz. 
The proposed capacitive sensor electronics system is shown in Fig.~\ref{fig.elec.1}. It is based on an AC differential charge amplifier~\citep{lotters1999} unlike the differential transformer designs used in other accelerometers~\citep{josselin1999,speake1997,weber2002}. 

The TM sensing bias is applied on opposite sides of the TM at $f_{0}$ = 100\,kHz using the injection electrodes, which induce an oscillating voltage in the TM, $V_{\rm TM}$, of about one tenth of the voltage applied to the electrodes~\citep{mance2012}. The two charge amplifiers at opposite faces of the TM sense the capacitances between the TM and the sensing electrodes. The outputs for each arm are (at 100\,kHz) 
$V_{i}=\frac{C_{i}}{C_{\rm f}}V_{\rm TM}$ ($i$=1,2)
where $C_{\rm f}$ is the feedback capacitor in the charge amplifier that defines the gain and using a parallel plate approximation, 
$C_{i} = \varepsilon_{0}\frac{A}{d\pm x}$ 
where $A$, $d$ and $x$ are the area of the sensing electrode, the gap between the TM and the electrode, and TM motion, respectively. The $\pm$ symbol indicates that the change in the capacitance is anti-symmetric: TM motion in one direction causes an increase in $C_{1}$ and a decrease in $C_{2}$ and vice versa. The small changes in the TM position are encoded in the differential signal, which is obtained using an instrumentation amplifier (IA). The output of the IA is band-pass filtered with a center frequency of 100\,kHz and sampled with an analog-to-digital converter (ADC). The amplitude and phase of the 100\,kHz signal is obtained after I/Q demodulation. The amplitude is proportional to the TM motion
\begin{equation}
    \Delta V\simeq2\frac{GV_{\rm TM}}{C_{\rm f}}\frac{C_{0}}{d}\Delta x
\end{equation}
where $G=40$ includes the gain of the IA and of the band-pass filter at 100\,kHz. $C_{0}$ is the capacitance between the TM and the electrode for $x$ = 0. 

The most critical parts of the electronics are the charge amplifiers. For the values given in Fig.~\ref{fig.elec.1}, the noise levels at 100\,kHz referred to the input of the IA are 51\,nV/Hz$^{1/2}$ and are dominated by the charge amplifier stages (IA, BPF and ADC noise levels are negligible). The equivalent position sensing noise is readily calculated by considering the position sensitivity
\begin{equation}\label{eq.sens}
\frac{1}{G}\frac{{\rm d}\Delta V}{{\rm d}x}\simeq2\frac{C_{0}}{d}\frac{V_{\rm TM}}{C_{\rm f}}
\end{equation}
For $C_{0}$ = 3.9\,pF (capacitance in the most sensitive axis, $x$-direction), $d$ = 1\,mm, $V_{\rm TM}$ = 0.5\,V and $C_{\rm f}$ = 10\,pF, it is 390\,V/m, which yields a noise equivalent position of 0.13\,nm/${\rm Hz}^{1/2}$, i.e. well below the 1\,nm/${\rm Hz}^{1/2}$ goal. 
A similar analysis for the angular readout system results in an angular sensitivity of $\sim$50\,nrad/Hz$^{1/2}$. 

Another source of error is due to common mode rejection (CMR) limitations and, more importantly, their fluctuations at the mHz frequency due to temperature variations. In this case, CMR fluctuations are dominated by the gain mismatch between the two charge amplifiers. The induced error in the TM position is $\delta x = d\alpha_{C_{\rm f}} \delta T$ where $\alpha_{C_{\rm f}}$ is the temperature coefficient of $C_{\rm f}$ and $\delta T$ is the temperature fluctuation. This implies that the temperature coefficient has to be at the ppm/K level for 1\,K/${\rm Hz}^{1/2}$ temperature stability. 

\begin{figure}[h!]
\centering
\includegraphics[width=\linewidth]{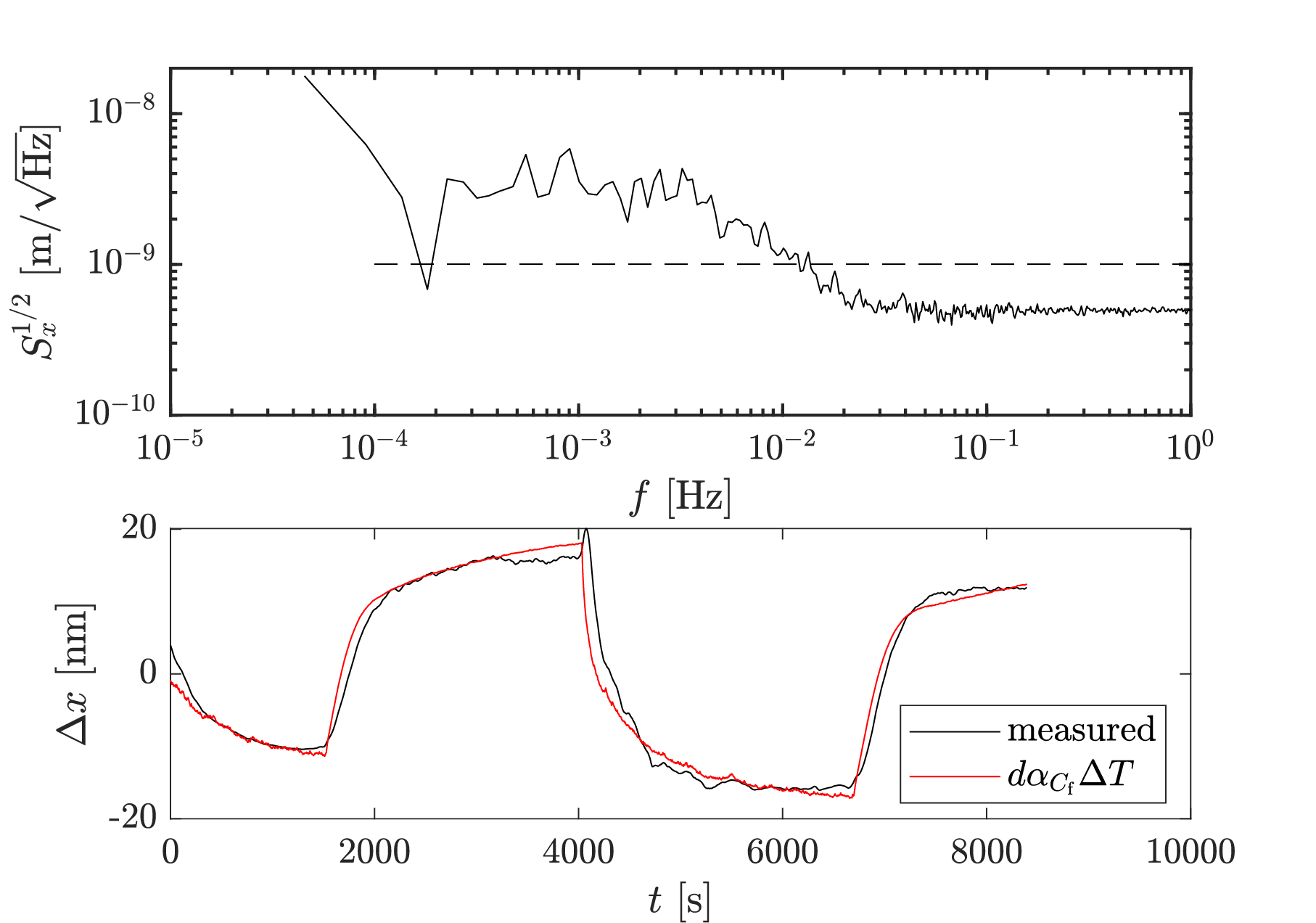}
\caption{Top: sensing electronics noise in m/Hz$^{1/2}$ (assuming $V_{\rm TM}$ = 0.5\,V). Bottom: measured temperature coefficient. The best fit value is 3.8\,nm/K. Both measurements are taken with fixed capacitors simulating the TM capacitances.
 \label{fig.elec.2}}
\end{figure}

The experimental results for a single degree-of-freedom electronics prototype are shown in Fig.~\ref{fig.elec.2}. The upper plot shows the ASD for a 6-hour measurement. The conversion to position noise has been done assuming $V_{\rm TM}$ = 0.5\,V. The noise levels for $f>20$\,mHz are 0.5\,nm/Hz$^{1/2}$, i.e., a factor $\sim$4 higher than the theoretical value, yet below 1\,nm/Hz$^{1/2}$. For $f<$20\,mHz, the noise levels increase due to temperature fluctuations and exhibit a plateau at 3\,nm/${\rm Hz}^{1/2}$ between 0.1\,mHz and 3\,mHz. The plot on the bottom shows the equivalent position changes when injecting temperature changes. The temperature coefficient is 3.8\,nm/K (3.8\,ppm/K for $d$ = 1\,mm). The temperature sensitivity of the charge amplifiers can be further reduced by selecting matching capacitors $C_{\rm f}$.

\subsection{Electrostatic actuation}

Electrostatic actuation is needed to keep the TM centered in the EH. The performance of such actuation is especially important in non-drag compensated mode since the applied electrostatic force is the signal used to derive the non-inertial forces acting on the spacecraft. The electrostatic forces are applied to the sensing electrodes at lower frequencies (50-300 Hz) to avoid interfering with the 100 kHz sensing signal.
The circuit for the actuating signals is rather simple---see Fig.~\ref{fig.elec.1}: a digital-to-analog converter (DAC) with a high-stability voltage reference and an active band-pass filter with gain. However, the requirements are stringent. The force on the TM for a single electrode is
 $   F\simeq \frac{1}{2}\frac{C_{0}}{d}V_{\rm act}^{2}$
where $V_{\rm act}$ is the actuation signal at $\sim$100\,Hz. In acceleration mode, 
the electrostatic force on the TM must be equal to the drag on the spacecraft to avoid a collision between them. At an altitude of 500\,km, the drag on a GRACE-like spacecraft is about 
0.3\,$\mu$m/s$^{2}$~\citep{mehta2017}, which requires an applied voltage of $V_{\rm act}\simeq$12\,V with relative stability requirements at the mHz band and at $2f_{\rm act}$ below $10^{-6}$\,1/${\rm Hz}^{1/2}$. At $f_{\rm act}$ the requirement is about 20\,$\mu$V/${\rm Hz}^{1/2}$. The stability at $f_{\rm act}$ and $2f_{\rm act}$ is not a problem for the actuator circuit. However, the requirement in the mHz band is challenging and, at a minimum, depends on the quality of the voltage reference of the DAC. Ultra-stable voltage references are available at the $10^{-6}$~1/${\rm Hz}^{1/2}$ level~\citep{halloin2013}. The requirement at lower altitudes, e.g., 350\,km is more demanding~\citep{montenbruck2000}, which means $10^{-12}$\,m/s$^{2}/{\rm Hz}^{1/2}$ noise levels cannot be reached unless a drag-free scheme is considered.

\section{Acceleration Noise Performance}
\label{sec:performance}

A detailed acceleration noise model for the S-GRS has been developed based on two operational scenarios: non-drag-compensated at an orbit altitude of 500 km (similar to GRACE and GRACE-FO), and drag-compensated at an altitude of 350\,km. 
The former and latter cases represent the worst and best realistic operational scenarios in terms of gravity recovery sensitivity. The model accounts for roughly 30 different noise sources that are relevant at acceleration noise levels below 
$1$\,pm/s$^{2}/{\rm Hz}^{1/2}$~\citep{gerardi2014}. This model is based on both torsion pendulum experiments, flight experiments performed on the LPF mission~\citep{armano2018} and measurements of the GRACE environment in orbit. The UF torsion pendulum, equipped with a LISA-like GRS, shown in Figure \ref{fig:EH_CMS}, has a demonstrated force noise performance below $1\,\rm{pN/Hz}^{1/2}$ at around 1\,mHz acting on lightweight test mass in one degree-of-freedom~\citep{ciani2017}. 

\begin{figure}
  \includegraphics[scale=0.6]{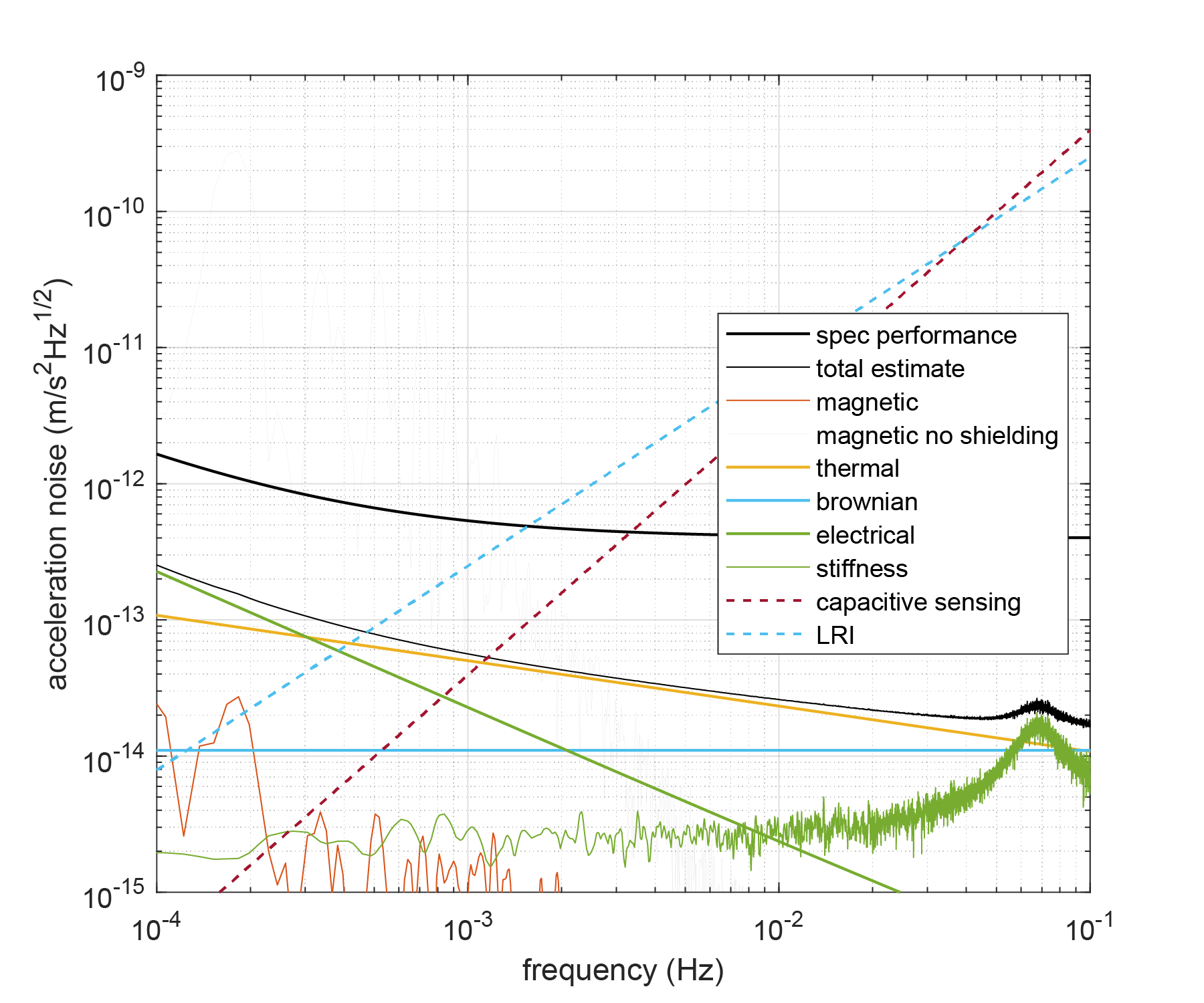}
  \includegraphics[scale=0.6]{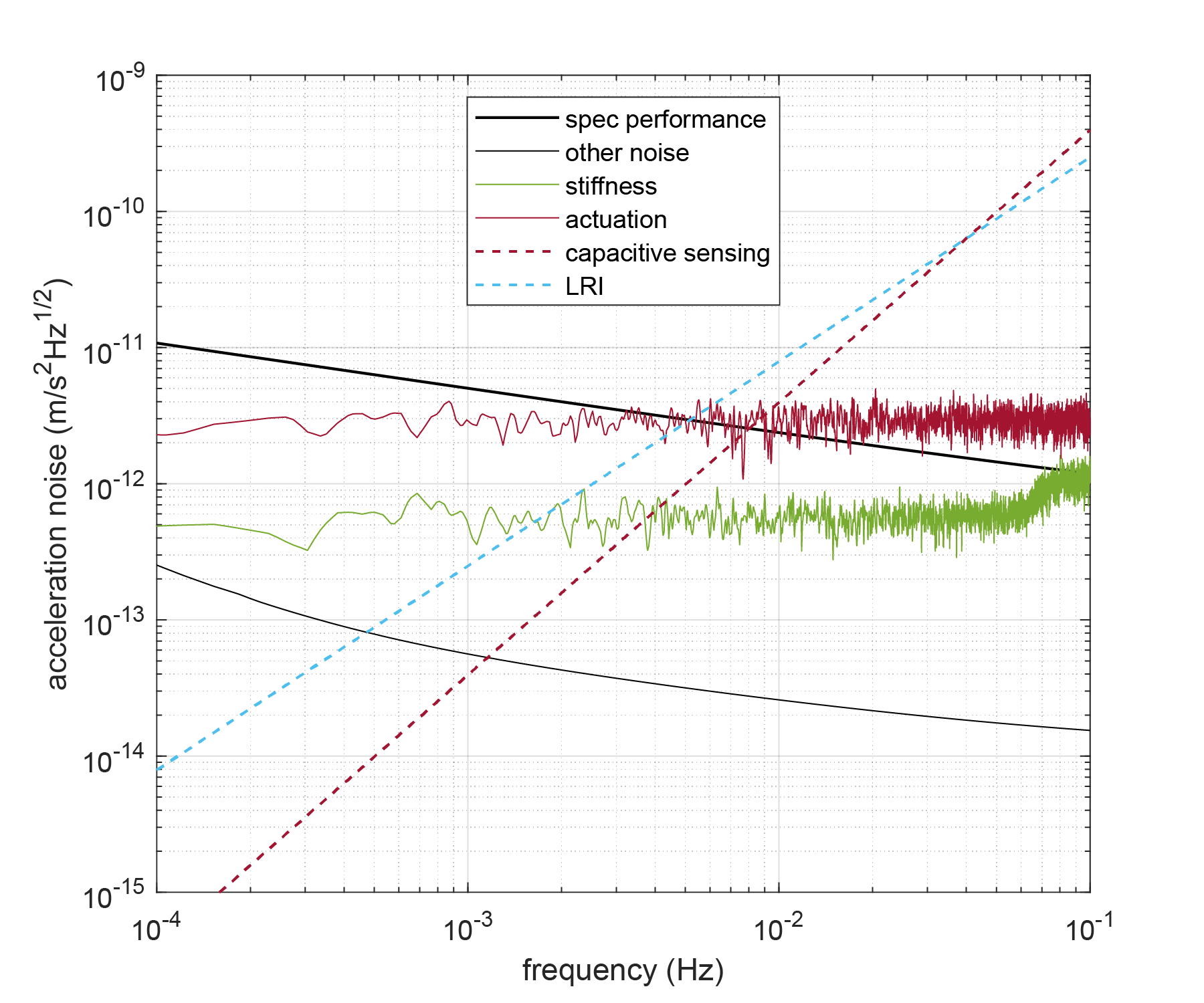}
\caption{Acceleration noise estimate for the simplified GRS,
operated on a drag-compensated platform at 350\,km altitude (top) and operated on a GRACE-like spacecraft at 500\,km (bottom), based on the measured GRACE-FO flight acceleration environment. See equations \ref{dragfree_asd} and \ref{nondrag_asd} for details.}
\label{fig:accelNoise}       
\end{figure}

Using the developed acceleration noise models and a spacecraft environment equivalent to that of GRACE-FO, Figure \ref{fig:accelNoise}(a) shows the acceleration noise performance of the simplified GRS if operated on a drag-free platform at 350\,km. Figure \ref{fig:accelNoise}(b) shows the equivalent performance in a GRACE-like mission at 500\,km with no drag-compensation. 
The performance of the S-GRS is expected to be roughly 40 times better than the GRACE accelerometers in the same mission configuration. If the sensor were operated on a drag-compensated platform, its performance would improve by an additional factor of 8 to about $7\times10^{-13}$\,m/s$^{2}$/Hz$^{1/2}$ around 1\,mHz.

The ASD of the acceleration noise, $S^{1/2}_a$, in the drag-free environment can be described by the function of frequency in Eq.~\ref{dragfree_asd}. For the non-drag-compensated 500\,km environment, the equivalent function describing the acceleration noise performance is seen in Eq.~\ref{nondrag_asd}. 

\begin{eqnarray}
\label{dragfree_asd}
  S_a^{1/2} & = & 4 \times 10^{-13} \mathrm{\,m/s^2/Hz^{1/2}} \nonumber \\ 
  && \sqrt{1 + \left( \frac{700 \mathrm{\,\mu Hz}}{f} \right) + \left( \frac{300 \mathrm{\,\mu Hz}}{f} \right)^2}.
\end{eqnarray}

\begin{equation}
\label{nondrag_asd}
  S_a^{1/2} = 5 \times10^{-13} \sqrt{1 + \left(\frac{1 \mathrm{\,Hz}}{f} \right)^{2/3}}\, \mathrm{\,m/s^2/Hz^{1/2}}
\end{equation}

These two ASDs are shown in bold in Figure \ref{fig:accelNoise} and indicate the acceleration noise budget for the S-GRS in the two orbital environments. These curves have been used to evaluate the gravity recovery capability of the S-GRS in Section \ref{sec:performance} over the frequency band of 0.1\,mHz up to 0.1\,Hz. 

The individual terms for the S-GRS noise model are grouped into six categories: stiffness, magnetic, thermal, Brownian, electrostatic, and actuation~\citep{gerardi2014}. 
Stiffness describes the coupling between the relative motion of spacecraft and TM and force gradients to produce an acceleration.
Magnetic forces arise through the interaction of the bulk material of the TM with the magnetic field at the TM, comprising contributions from the Earth field and spacecraft components.
Fluctuating temperatures and thermal gradients in the S-GRS create TM acceleration noise, through residual gas, radiation pressure and outgassing. 
Brownian noise is caused by random impacts of the residual gas molecules inside the vacuum chamber on the TM.
Electrostatic forces produced by the variation of surface potentials and charge build-up on and around the TM create unwanted accelerations.
Finally, noisy forces needed to control the TM position contribute to the total acceleration noise either directly in the sensitive axis or by cross-talk from orthogonal degrees-of-freedom.
A calculation of each of these noise sources is included in Figure \ref{fig:accelNoise}, as colored traces and the best estimate of the total acceleration noise is shown by the thin black curve. Actuation noise is not included in Figure \ref{fig:accelNoise}(a) because test mass actuation along the sensitive axis of the instrument is not required in a drag free configuration. In Figure \ref{fig:accelNoise}(b), the combined thermal, magnetic, Brownian and electrical effects are identical to the drag free case and are shown as the gray curve. The actuation noise and stiffness contributions in this mode are significantly higher because of the need to control the test mass position electrostatically and are therefore plotted explicitly. 
Also shown in Figures \ref{fig:accelNoise}(a) and \ref{fig:accelNoise}(b) are the displacement measurement noise curves for the LRI and S-GRS capacitive readout, twice differentiated to convert them to acceleration noise. Note that since LRI and S-GRS noise spectra are quite similar, neither one dominates the overall gravity recovery sensitivity.

\subsection{Detailed Noise Simulations}


In the following sections we summarize the calculations of the major noise sources contributing to the acceleration noise budget of the S-GRS.

\paragraph{Thermal Effects}
Thermal acceleration noise effects in the S-GRS are driven by the fluctuation in temperature gradients across the TM. There are three physical sources of thermal gradient-induced noise: the radiometer effect, in which the residual gas in the gaps between the TM and EH has different pressure on opposite sides of the TM. Radiation pressure caused by the difference in energy of photons emitted by the hot and cold sides of the TM and EH, and the difference in temperature-dependent outgassing rates between hot and cold surfaces. Radiation pressure is well described by analytical and geometrical expressions adapted from those derived for LISA and LPF~\citep{carbone2007}. 
Both radiometer and outgassing effects depend on the geometry-dependent gas conductances of the EH. Because of this, the effects are simulated with the numerical analysis software MolFlow~\citep{kersevan2009}. Using a simplified geometry of the TM and inner EH, the force per unit temperature gradient is determined by applying temperatures to the EH faces and calculating the pressure difference on the TM faces. Assuming water is the dominant outgassing species, an operational temperature of 293\,K and a residual gas pressure of $10^{-5}$\,Pa, temperature-gradient coupling parameters can be calculated for all three effects.
To calculate the expected TM acceleration noise assumptions must be made about the stability of the temperature gradient across the EH. Data from the GRACE-FO mission, show that the temperature gradient stability at the accelerometer instrument is better than 5\,mK/Hz$^{1/2}$ at 1\,mHz. Assuming a performance at this level for the S-GRS gives an upper limit of 0.05\,pm/s$^2$/Hz$^{1/2}$ for thermal effects at 1\,mHz. A frequency dependence of $f^{-1/2}$ is assumed. 





\paragraph{Gravitational Effects}
The mass distribution on-board the spacecraft will produce a residual gravitational force and force gradient at the TM. 
The resulting acceleration of the TM is likely to be small compared to the drag on the spacecraft so that in non-drag free mode the acceleration is a small fraction of the actuation force. In a drag compensated mode of operation however, a continuous acceleration of the TM along the line of sight between spacecraft may eventually affect the inter-spacecraft laser metrology and require corrective maneuvers. This constraint will enforce some level of gravitational balancing on the spacecraft system. The LISA Pathfinder mission demonstrated the ability to compensate TM accelerations at the sub-pm/s$^2$ level~\citep{ferroni_selfgravity_2016} but S-GRS performance likely does not require such fidelity. 

In order to estimate the residual gravitational force on the TM, a mapping of components from a geometrical model of the GRACE-FO spacecraft was analyzed to estimate the volume and center of mass of each component, the S-GRS CAD design was also broken down into 127 mass elements. 
The gravitational force on the TM due to each component and S-GRS mass element was calculated resulting in a net residual dc acceleration on the order of 30\,nm/s$^2$. This level of acceleration would result in an accumulated length change in the inter spacecraft distance of around 10\,km in 10 days but could be mitigated by better gravitational balancing, station-keeping, or compensation by electrostatic actuation of the TM. The gravitational force also produces a force gradient at the TM of 13.5\,nN/m in the initial estimate. This is an important contribution to the total stiffness of the TM. In band fluctuations of the gravitational force due to thermal distortion of the spacecraft were calculated and found to be negligible. This preliminary calculation provides only an order of magnitude estimation of the gravitational effects given that a flight platform for the S-GRS has not been defined. 

\paragraph{Magnetic Effects}
The magnetic force $F_{B,x}$ on the TM along the free-falling $x$-axis can be written as
\begin{equation}\label{eq:mag}
    F_{B,x} = \left( \frac{\chi_{\mathrm{TM}}L^{3}_{\mathrm{TM}}}{\mu_0}\vec{B}+\vec{M}\right) \cdot \nabla B_{x}, 
\end{equation}
where $\vec{M}$ is the remnant magnetization of the TM, $\chi_{\mathrm{TM}}$ is the magnetic susceptibility, 
$L_{\mathrm{TM}}$ is the length of the TM side, and 
$\vec{B}$ is the magnetic field at the TM. The remnant magnetization $M$ of the gold platinum TM is expected to be around 10$^{-9}$\,A\,m$^2$, while the susceptibility $\chi$ is around 10$^{-5}$. The magnetic field at the TM originates from components on-board the spacecraft, and from Earth. The field gradient is dominated by the on-board component. 
The time-varying field at the spacecraft due to its motion through the Earth field, coupled with the static local field gradient gives rise to a magnetic force signal at harmonics of the orbital frequency. Stochastic noise in $F_{B,x}$ arises from the coupling between the noise in the field gradient due to spacecraft equipment and the rms Earth field.

In order to estimate the magnitude of the magnetic effects, a magnetic map of the spacecraft has been created, similar to the gravitational analysis described above.
Each spacecraft component is represented as a magnetic dipole at a position determined from a spacecraft geometrical model. The magnitude of the dipole is assigned to each element of the spacecraft equipment based on published measurements made on similar components of the LPF mission~\citep{martin2015}. 
A simulation was developed, assigning each dipole a uniformly distributed random direction and calculating the field and gradient at the TM. 10,000 field realizations were calculated, which resulted in a local field magnitude of $\pm 0.5\,\mu$T and a gradient of $\pm 2.7 \,\mu$T/m (1-$\sigma$). This is likely to be an optimistic estimate based on the simplicity of the geometry used. Given that the LISA Pathfinder gradient requirement was 5\,$\mu$T/m with stringent magnetic cleanliness requirements we assume a worst case gradient of 500\,$\mu$T/m in our acceleration noise calculations.
The Earth field is measured on-board by GRACE-FO. The along track component of the field has a peak-to-peak amplitude of around 55\,$\mu$T. Combining these assumptions in Equation (\ref{eq:mag}) results in the faint curve in Figure \ref{fig:accelNoise}(a) which dominates the acceleration noise budget and motivates the inclusion of a magnetic shield within the S-GRS.





Initial estimates using a realistic geometry simulated in the COMSOL simulation toolkit indicate the magnetic shield will reduce the field and gradient at the TM by up to a factor 100 leading to an acceleration noise contribution represented by the red curve in Figure \ref{fig:accelNoise}(a), suppressed by a factor 10$^4$ compared to the unshielded case. Incorporating a magnetic shield therefore relaxes significantly the magnetic cleanliness requirements compared to LISA Pathfinder making it easier to accommodate the  S-GRS as a secondary payload or on a spacecraft requiring an electrical thruster for drag compensation. 

\paragraph{Other acceleration noise sources}

Electrostatic interactions between the charged TM and electric fields in the EH result in forces on the TM. Both the stray electric field and the charge build up on the TM are noisy resulting in an acceleration noise. The effect has been well studied in the context of the LISA mission~\citep{Sumner2020}. Applying geometric scaling factors to those model predictions to account for the reduction in size of the S-GRS TM and gaps, and assuming a well-performing CMS, we expect a noise of around 20\,fm/s$^{2}$/Hz$^{1/2}$. This level assumes continuous TM charge control using photoemission from UV light. In this case, the acceleration noise is dominated by the discharging current shot noise with a $1/f$ shape and a magnitude that is expected to be similar to that expected for LISA. Assuming a worst case of uncompensated stray electric fields in the S-GRS leads to the green curve of Figure \ref{fig:accelNoise}(a).

Electrostatic actuation forces applied to control the TM will introduce force noise through the imperfect stability of the voltages applied to actuation electrodes in a non drag-compensated mode of operation. The expected relative voltage stability for actuation voltages is around $5\times10^{-6}/Hz^{1/2}$ resulting in a force noise with a spectral density at a level of around $10^{-5}$ times the applied actuation force. The magnitude of this effect is calculated from the control and actuation simulations described in Section \ref{sec:Controls} below. In a non-drag compensated operational mode, this is a limiting contributor to the S-GRS acceleration noise performance shown by the dark red line in Figure \ref{fig:accelNoise}(b).

As well as force noise, the electrostatic force needed to control the test mass position  creates a significant force gradient. From the electrostatic control simulations, the commanded voltages on the electrodes produce a force gradient of $3\times 10^{-4}$\,N/m, while for a drag-compensated mode, the effect is reduced to around $10^{-6}$\,N/m, contributed by the gravitational forces and the electrostatic sensing bias of the TM. Coupling these gradient values with the relative motion of the TM and SC shown in Figure \ref{1dof.pos} produces the light green traces of Figure \ref{fig:accelNoise}(a) and (b). 

\subsection{Mission Architecture Considerations}

Like previous Earth geodesy missions (and space gravitational wave interferometers), the proposed concept that includes the S-GRS, LRI, GPS receiver, and spacecraft forms a single “instrument”. The performance of this instrument is directly tied to its attitude and possible drag-compensation control performance; as well as the thermal, electromagnetic, and gravitational environment provided for the S-GRS and LRI. Therefore, to maximize scientific performance, both the instrument and spacecraft platform must be considered. A spacecraft environment equivalent to that of GRACE-FO is assumed for the performance model described in the previous section.  
While the GRACE spacecraft busses provided a reasonably benign environment in low Earth orbit, no extreme measures were taken to control temperature, gravity and electromagnetics, as for example on the LPF spacecraft. Nevertheless, the S-GRS in the GRACE environment provides significant improvement. The mitigation steps taken in the S-GRS design include a small mu-metal magnetic shield surrounding the EH and the vacuum enclosure (see Figure \ref{fig:S-GRS_Head}). Including these measures, the key environmental parameters needed to achieve the stated performance are listed in Table \ref{tab:envParams}.

\begin{table}
\caption{Required environmental parameters for the S-GRS}
\label{tab:envParams}       
\begin{tabular}{ll}
\hline\noalign{\smallskip}
Quantity & Value  \\
\noalign{\smallskip}\hline\noalign{\smallskip}
S/C-to-TM jitter (drag-comp.) & 50 $\mathrm{nm/Hz^{1/2}}$ \\
S/C-to-TM jitter (non-drag-comp.) & 2 $\mathrm{\mu m/Hz^{1/2}}$ \\
Magnetic field at the TM & 100 $\mathrm{\mu T}$ \\
Magnetic field fluctuation at the TM & 2 $\mathrm{\mu T/Hz^{1/2}}$ \\
Magnetic field gradient at the TM & 20 $\mathrm{\mu T/m}$ \\
Mag. field gradient fluctuation at TM & 0.5 $\mathrm{\mu T/m Hz^{1/2}}$ \\
Mean temperature at the TM & 293 $\mathrm{K}$ \\
Temperature fluctuations across EH & 5 $\mathrm{mK/Hz^{1/2}}$ \\
Temperature fluctuations near TM & 1 $\mathrm{K/Hz^{1/2}}$ \\
Pressure around the TM & 10 $\mathrm{\mu Pa}$ \\
\noalign{\smallskip}\hline
\end{tabular}
\end{table}

The performance of the instrument in terms of gravity recovery accuracy is also closely tied to orbit selection and the number of orbits (satellite pairs). The number of pairs of satellites that is financially viable depends on the launch mass of each spacecraft and therefore the size, weight and power of the S-GRS and LRI. Finally, mission lifetime and orbit altitude also depend on whether a drag-compensated platform is chosen.

The estimated mass of the S-GRS Head and electronics unit is $\sim$8 kg and $\sim$5 kg respectively. The power consumption of the S-GRS electronics unit is $\sim$20 W. The total mass of the GRACE-FO LRI units per spacecraft is 25\,kg, and the nominal power consumption is 35 W. The sum of these results is a mass of $\sim$38 kg and power consumption of $\sim$55 W. These values already put the mass and power budget of the S-GRS and LRI within levels typically accommodated on ESPA-class micro-satellites, whose size and mass are 1/3 of the GRACE-FO spacecraft.

\section{S-GRS and Drag-compensation Control}\label{sec:Controls}


In both drag compensated and non-drag compensated modes of operation, the TM position must be controlled, either by actuation of the spacecraft with micro-Newton thrusters or by electrostatic actuation of the TM. The residual jitter of the TM position relative to the spacecraft drives the stiffness contribution to acceleration noise budget and, in the non-drag compensated case, the noisy TM actuation force is an important noise contribution. We have developed a control system for both modes of operation and here we show the performance in the critical $x$ (SC-SC) direction.

The performance is evaluated through a numerical simulation that uses realistic sensing noise, actuation noise, and spacecraft environment. The control of the spacecraft is simulated in both the non-drag-compensated and drag-compensated case. 
In both cases, a state space representation of a simple equation of motion for the linear satellite dynamics is used.
\begin{equation}
    \ddot{x} = a_{\mathrm{SC}} + a_{\mathrm{TM}} + \frac{K}{m_{\mathrm{TM}}}x
\end{equation}
where $a_{\mathrm{SC}}$ is the environmental disturbance on the spacecraft, $a_{\mathrm{TM}}$ is the direct TM force noise, $K$ is the stiffness, $m_{\mathrm{TM}}$ is the mass of the TM and $x$ is the relative position of the TM with respect to the SC. The position sensor noise and applied actuation forces are accounted for when applying this equation in the simulation.

The controls simulation is done using MATLAB and Simulink to simulate around one day of data. For the non-drag compensated case at 500\,km altitude, the spacecraft acceleration noise is taken from previously recorded GRACE-FO Level-1B accelerometer data along the $x$-direction, which measures the relative acceleration between the test mass and spacecraft and we assume this measurement is dominated by the spacecraft acceleration signal. 
At 350\,km we assume the along-track disturbance is 10 times greater.
The position sensor noise in both cases is modeled as a Gaussian white noise with a level of 1\,nm/Hz$^{1/2}$ while the direct forces on the test mass, excluding actuation are approximated by a white noise of 0.2\,pN/Hz$^{1/2}$ . A spring constant of 1$\mu$\,N/m, is used for the drag-compensated case in which electrostatic stiffness from TM actuation forces will be small, in a non-drag compensated case where large voltages are applied to sensor electrodes, a value of 310\,$\mu$N/m is used. This figure was calculated from the commanded voltages output from the simulation described below, iterating from an initial value of 100\,$\mu$N/m for the stiffness.

\begin{figure}
\centering
\includegraphics[scale=0.17]{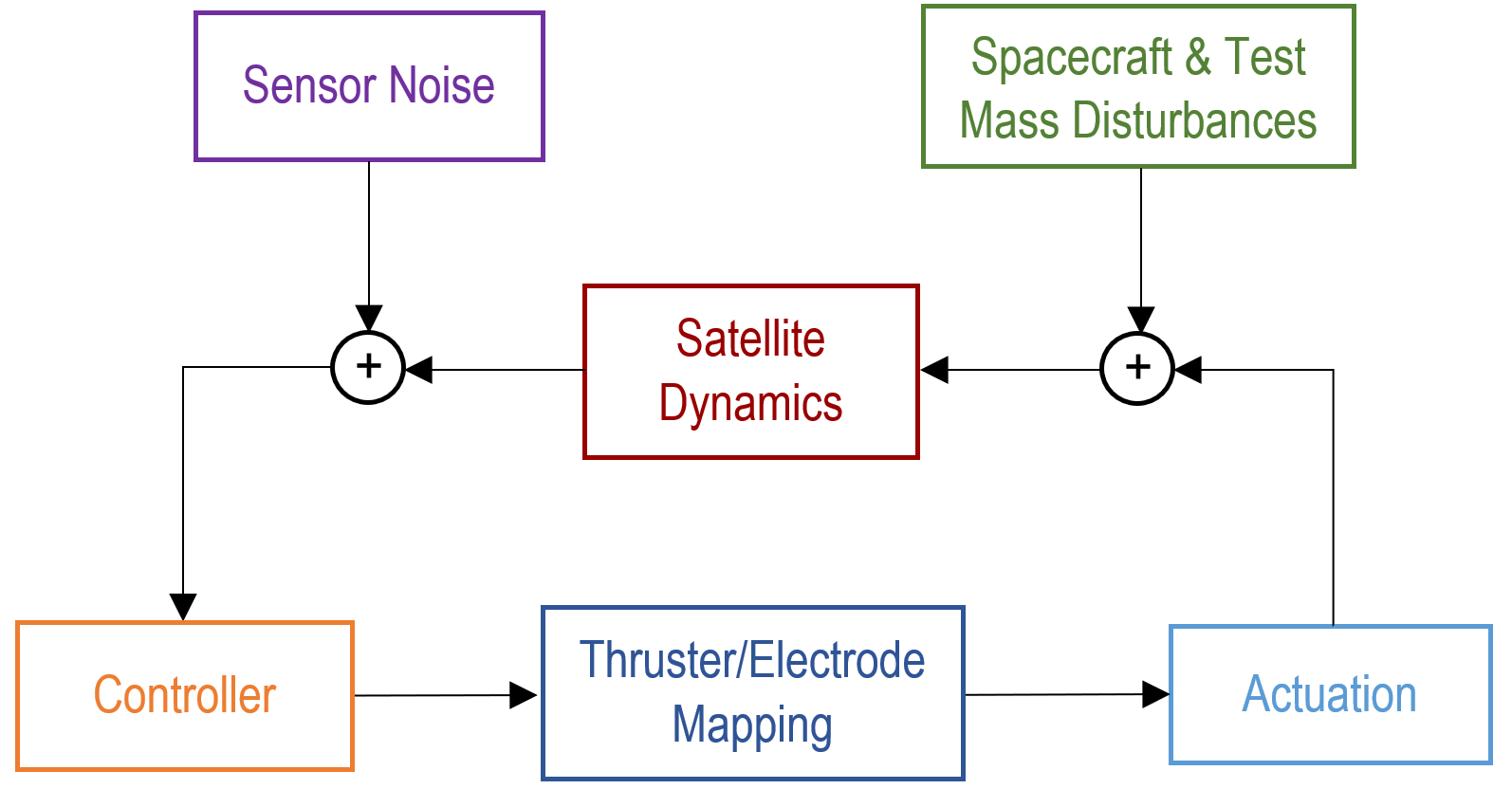}
\caption{Control loop block diagram.
\label{fig:control}}
\end{figure}

The feedback control block diagram is shown in Figure \ref{fig:control} indicating the modelled disturbances, sensing and actuation noises. 
In order to reduce stiffness contributions to the acceleration noise budget, the aim of the controller in both control schemes is to reduce the in-band motion of the test mass to below 50\,nm/Hz$^{1/2}$ for drag-compensated and below 2\,$\mu$m/Hz$^{1/2}$ for non-drag-compensated. 
In both cases, an H-infinity optimization method is employed to determine the control effort needed to stabilize the motion. This method allows for the disturbances and noise to be accounted for in the control algorithm by using a Multi-Input Multi-Output system and indicating which error outputs are to be kept small. Using the function {\tt hinfsyn} in MATLAB, we minimized the position readout and velocity jitter.

The thruster model for the drag-compensated simulation is based on cold-gas thrusters used in the Gaia, Microscope and LISA Pathfinder missions~\citep{LPFColdGas}. Our model is capable of simulating 6-dof SC control although here we present only the results for the critical $x$ direction. Similar to the Grace thruster system, four thrusters actuate each degree of freedom, ($x, y, z$), linear thrust is achieved by actuating thrusters on the same face of the spacecraft, torques are obtained commanding thrust to opposite pairs. 
The thrusters are positioned so that translational thrust produces no torque and torque applied to the spacecraft produces no net force. Once the command to each individual thruster has been calculated, the constraints of a cold gas thruster model are applied. A resolution of 0.1\,$\mu$N and range of 2 to 1000\,$\mu$N is used, with a time response of 250 ms to 63\% of the commanded thrust level. A thrust noise of 0.17\,$\mu$N/Hz$^{1/2}$ is also added in this step. The maximum thrust used here is extended beyond the capability of the system included on LISA Pathfinder. It  remains to be investigated whether there are technical barriers to achieving this thrust level, however, a number of other micro-thruster technologies are under development which may address the required thrust range and resolution.


For the non-drag-compensated case, the S-GRS electrodes are utilized as actuators on the test mass. Commanded forces and torques are converted to electrode voltages using a simple electrostatic model. Assuming a maximum applied voltage of 15\,V with 16-bits of resolution, the maximum force that can be applied to the TM in our model is about 0.4\,$\mu$N with resolution of 0.1\,fN. The ASD of the voltage noise is calculated as 5\,ppm of the rms commanded voltage on each electrode. This noise is added to the commanded force before calculating the resulting force on the test mass. 

\begin{figure}
  \includegraphics[scale=0.63]{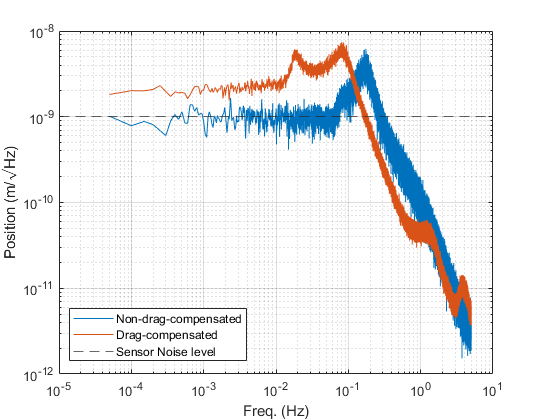}
  \caption{One degree-of-freedom in-loop position ASD for drag-compensated and non-drag-compensated cases.}
  \label{1dof.pos}
\end{figure}

The results of the $x$-axis position simulation for drag-compensated and non-drag compensated cases are shown in Figure~\ref{1dof.pos}. The drag-compensated simulation results show that the larger input noise from the 350\,km orbit is causing this case to be gain limited, but still within the 50\,nm/Hz$^{1/2}$ jitter requirements. The non-drag-compensated simulation shows that the in-loop position measurement is at a level below the measurement noise successfully suppressing stiffness contributions to the acceleration noise budget. The controller performs optimally so that the sensor noise level is the limiting factor and there is no wasted control effort. They both make a negligible contribution to the acceleration noise budget.

\begin{figure}
  \includegraphics[scale=0.56]{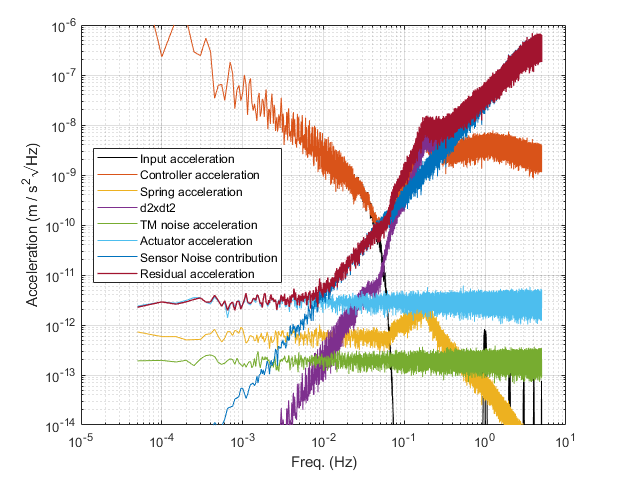}
  \raggedleft
  \caption{Residual acceleration for the non-drag-compensated case.}
  \label{1dof.ndf.resid}
\end{figure}

Further analyzing the non-drag compensated case, we can confirm the limiting sensitivity of the S-GRS from the simulated data. In this mode of operation the spacecraft acceleration would be determined from the S-GRS measurements and the applied control forces. Figure~\ref{1dof.ndf.resid} shows the ASD of various sources of apparent TM acceleration in the simulation. The red curve shows the input SC acceleration to the simulation. After subtracting the commanded control forces, the residual acceleration is given by the dark red curve. This represents the performance limit of the S-GRS at $3 \times 10^{-12}$\, m/s$^2$Hz$^{1/2}$ and is limited by the actuation noise (light blue curve) and position sensor noise (dark blue curve). Also shown are the sub-dominant contributions of stiffness (yellow), direct TM forces (green) and in-loop TM motion estimate (purple).



%

\section{Conclusions and Outlook}
We have developed a novel simplified gravitational reference sensor optimized for future Earth geodesy missions. This sensor is a scaled-down version of the flight LPF GRS with reduced complexity. The performance is roughly 40 times better than that of the accelerometers flown on GRACE and GRACE-FO when operated on a non-drag-compensated spacecraft in a 500 km altitude Earth orbit. When operated on a drag-compensated spacecraft the performance improvement is a factor of 500. This is achieved by eliminating the small grounding wire used in prior accelerometers and instead using a UV photoemission-based TM charge control scheme. In addition to eliminating the noise associated with the grounding wire, this also allows for a larger TM and a larger TM-to-electrode housing gap. We described the mechanical and electrical design of the S-GRS and provided details of our performance assessment, which is based on models validated by LPF and torsion pendulum experiments, and used flight environment data from GRACE-FO. Gravity recovery simulations reveal that the S-GRS noise and the GRACE-FO LRI noise contribute equally to the gravity recovery sensitivity of a mission that utilizes both. Therefore, neither the S-GRS nor the LRI would limit measurement sensitivity.

A team led by the University of Florida, in collaboration with Caltech/JPL and Ball Aerospace is advancing the technology readiness level of the S-GRS. Currently considered TRL 3-5, depending on the component, we plan to reach TRL 5-6 in three years and flight readiness by the mid-to-late 2020's.

\section*{A Appendix}
High fidelity numerical simulations relying on the same software suite used to process GRACE and GRACE-FO data at JPL were run to quantify performance in Figures \ref{fig:degreeVar} and \ref{fig:spatialScale}.  The simulations consist of a truth run to create a set of synthetic satellite observations based on a realistic flight environment, followed by a nominal run where the truth measurements are perturbed in some way.  The error in the recovered gravity field due to this perturbation is then quantified via a large linear least squares estimation process, as is commonly used in the GRACE and GRACE-FO data processing.  Force models used in the truth and nominal runs are given in Table \ref{tab:forcemodels} when temporal aliasing error is included in the simulation.  When only measurement system error is considered, the nominal models in Table \ref{tab:forcemodels} are set equivalent to the truth models, so there is no perturbation among the model.  The simulation timeframe is January 1-29, 2006.  The gravity estimation process is a 2-step process, where in the first step, a set of “local” parameters are estimated using the tracking data to converge to a best fitting orbit.  These parameters consist of daily position and velocity of each spacecraft, daily accelerometer scale factors and biases, and range-rate biases, drifts, and one cycle per revolution each orbital revolution.  In the second step of the gravity estimation process, these same parameters are again adjusted along with a 29-day mean gravity field expressed to spherical harmonic degree and order 180.

\begin{table}[h!]
\caption{Force Models used in Numerical Simulations}
\label{tab:forcemodels}
\begin{tabular}{ |p{2.4cm}|p{2.2cm}|p{2.2cm}|  }
\hline
Measurement & Truth Model & Nominal Model  \\
\hline
Static Gravity Field & gif48 & gif48 \\
\hline
Ocean Tides & GOT4.8 & FES2004\\
\hline
Nontidal Atmosphere and Ocean (AOD) & AOD RL05 & AOerr + DEAL \citep{dobslaw2016}\\
\hline
Hydrology + Ice & ESA Earth System Model & N/A\\
\hline
\end{tabular}
\end{table}

Measurement system errors ingested into the simulation process rely largely on heritage information from GRACE-FO.  Orbit error is introduced by adding white noise with a 1 cm standard deviation in all 3-axes with a 5-minute sampling time.  Attitude Error is derived using the difference of two competing data products used to define the GRACE-FO attitude.  For pitch and yaw, the difference between an attitude solution that combines star tracker and IMU data (as in the GRACE-FO v04 SCA1B data product) with star tracker and laser steering mirror data (from the LRI) is used to define the error \citep{goswami2021}.  Since the laser steering mirror is insensitive to roll variations, roll error is defined as the difference between an attitude solution that uses only star tracker data with one that uses both star tracker and IMU data.  LRI error is derived from GRACE-FO flight data \citep{abich2019}, where laser frequency noise dominates high frequencies and tilt-to-length coupling error dominates lower frequencies \citep{wegener2020}.  The LRI error curves with ‘improved alignment’ assume the tilt-to-length coupling error is driven to zero via improved alignments relative to the center of mass of the spacecraft.   The GRACE-FO accelerometer error is taken as the best estimate of performance prior to launch of GRACE-FO and is approximately a factor 3 better than the requirement discussed in \citep{GRACE-FO}. The S-GRS error is described in detail in Section~\ref{sec:performance}.

\section*{Acknowledgements}
This work was supported by the NASA Earth Science Technology Office (ESTO) grant 80NSSC20K0324.

We thank Peter Bender for his insights into the benefits of improved accelerometry for future GRACE-like missions.


\bibliographystyle{spbasic}      

\bibliography{IIP_paper_references}


%
%


\end{document}